\documentclass[12pt]{article}
\usepackage{amsmath,bm}
\usepackage{bm}
\usepackage{epsfig}
\usepackage{color}

\renewcommand{\thefootnote}{\fnsymbol{footnote}}

\begin{document}

\title{
\begin{flushright}
\begin{minipage}{0.2\linewidth}
\normalsize
WU-HEP-14-07 \\*[50pt]
\end{minipage}
\end{flushright}
{\Large \bf 
Moduli inflation in five-dimensional supergravity models
\\*[20pt]}}

\author{Hiroyuki~Abe\footnote{
E-mail address: abe@waseda.jp} \ and \ 
Hajime~Otsuka\footnote{
E-mail address: hajime.13.gologo@akane.waseda.jp
}\\*[20pt]
{\it \normalsize 
Department of Physics, Waseda University, 
Tokyo 169-8555, Japan} \\*[50pt]}

\date{
\centerline{\small \bf Abstract}
\begin{minipage}{0.9\linewidth}
\medskip 
\medskip 
\small
We propose a simple but effective mechanism to realize an 
inflationary early universe consistent with the observed 
WMAP, Planck and/or BICEP2 data, which would be incorporated 
in various supersymmetric models of elementary particles constructed in the 
(effective) five-dimensional spacetime.
In our scenario, the inflaton field is identified with one of 
the moduli appearing when the fifth direction is compactified, 
and a successful cosmological inflation without 
the so-called $\eta$ problem can be achieved by a very simple 
moduli stabilization potential. 
We also discuss the related particle cosmology during 
and (just) after the inflation, 
such as the (no) cosmological moduli problem. 
\end{minipage}
}

\begin{titlepage}
\maketitle
\thispagestyle{empty}
\clearpage
\tableofcontents
\thispagestyle{empty}
\end{titlepage}

\renewcommand{\thefootnote}{\arabic{footnote}}
\setcounter{footnote}{0}

\section{Introduction}
\label{sec:intro}
The cosmological inflation at the early universe is an 
attractive scenario which can solve the flatness and the 
horizon problems, and simultaneously explains 
the density perturbation of the initial universe. 
Recent data from the Planck satellite~\cite{Ade:2013zuv} 
show that the primordial non-Gaussianity in the Cosmic 
Microwave Background (CMB) fluctuations is small and 
the spectral index $n_s$ is less than $1$, and set an 
upper bound on the tensor-to-scalar ratio $r$. 
In addition to the Planck result, BICEP2 
experiments reported the lower bound on the ratio $r$ \cite{Ade:2014xna}.
Because our universe is isotropic, most inflation models 
assume a Lorentz scalar field called inflaton field 
(which does not transform under the four-dimensional 
Lorentz transformation of our universe) with very 
specific forms of its potential terms (even of its 
kinetic term) in order to be consistent with observations, 
because parameters in the inflaton potential (as well as 
the kinetic term) are severely constrained by the CMB data. 
From the theoretical point of view, we may have to identify 
some specific origin of such an inflaton scalar field itself, 
otherwise it is difficult to restrict these parameters. 

One of the origin could be a modulus field which 
appears as a zero-mode of extra-dimensional components 
in vector and/or tensor fields in higher-dimensional 
spacetime with the compactified extra-dimensions. 
Parameters in the modulus scalar potential are constrained 
by the higher-dimensional Lorentz as well as gauge invariance. 
Moduli fields are ubiquitous in the superstring/M theory, 
one of the promising candidates for a unified description 
of elementary particles and gravity, whose low energy 
effective theories are described by supergravity. 
The vacuum expectation values of closed (open) string 
moduli fields determine, e.g., the size and the shape of 
extra-dimensional space (the position of D-branes and 
Wilson-lines of the gauge potential induced on them) 
and so on, which accordingly determine phenomenological 
aspects of the effective theory around the vacuum. 

Therefore, it is important to study moduli inflation 
scenarios based on the full supergravity framework, where 
the local supersymmetry plays important roles to determine 
the precise form of moduli kinetic and potential terms 
as well as their couplings to matter fields. 
The five-dimensional (5D) supergravity with the 
compact fifth dimension, which has a full off-shell 
formulation~\cite{Zucker:1999ej,Kugo:2000af}
with a local superconformal symmetry, provides a simple 
but the attractive starting point for such a study. 
Bacause a way of dimensional reduction keeping the off-shell 
structure was proposed~\cite{Abe:2006eg} in 4D ${\cal N}=1$ 
superspace~\cite{Abe:2004ar,Paccetti:2004ri}, 
we can derive a four-dimensional (4D) effective action 
for moduli and matter fields systematically, which has 
a full 4D ${\cal N}=1$ local super(conformal) symmetry. 
Then, we can easily write down the on-shell action (not 
only in the Einstein frame but also in any other frame, 
if necessary) for analyzing the moduli inflation and the 
related particle cosmology. 

In this paper, we study the moduli inflation starting from 
the 5D off-shell supergravity compactified on orbifold 
$S^1/Z_2$ with two fixed points. In addition to $Z_2$-even 
vector and hypermultiplets which include multiplets of 
supersymmetric standard model in their zero-modes, we 
introduce $Z_2$-odd vector fields forming $Z_2$-odd 
vector multiplets, whose fifth components yield multiple 
moduli forming chiral multiplets in the 4D effective theory. 
Numerous particle physics models were proposed so far in 
such an orbifold framework, where the chirality of the 
observed quarks and leptons arises as a consequence of 
the orbifold structure, and there is a mechanism to 
localize matter (and even gravity) fields exponentially 
in the fifth dimension, which can be a source of the 
observed hierarchical structure of quark and lepton 
masses and mixings~\cite{ArkaniHamed:1999dc}
\footnote{See Ref.~\cite{Abe:2008an} for a realization 
of the realistic flavor structure in the framework of off-sell 
dimensional reduction.}
(and of the huge hierarchy between the weak and the 
Planck scales~\cite{Randall:1999ee}). 
A superpotential for such localized matter fields 
is allowed at the fixed points where the supersymmetry 
is reduced, which can be a source of moduli potential as well, 
in addition to some nonperturbative effects such as 
a gaugino condensation. We expect the exponential 
form of the localized wavefunctions plays a certain 
role to realize a successful moduli inflation 
as well as their stabilization. 

This paper is organized as follows. 
In Sec.~\ref{sec:moduli}, we review the moduli fields 
appearing in 5D supergravity on $S^1/Z_2$. Then, 
we propose two simple models, one of them realizes 
the small-field inflation and the other does the 
large-field one in Secs.~\ref{sec:sinflation} and \ref{sec:linflation},  respectively, both are 
triggered by the moduli dynamics. 
We show their consistency with the recent 
observations. Sec.~\ref{sec:conclusion} 
is devoted to the conclusion. We show the 
canonical normalization of fields for the large-field 
model in Appendix~\ref{app:can}.

\section{Moduli effective action on orbifold $S^1/Z_2$}
\label{sec:moduli}
First in this section we review moduli effective action 
appearing from a compactification of 5D (off-shell) 
supergravity on orbifold $S^1/Z_2$. 
The most general form of the 5D background metric 
preserving a 4D flatness is given by 
$ds^2=G_{MN}dx^Mdx^N
=e^{-2\sigma(y)}\eta_{\mu \nu}dx^\mu dx^\nu -dy^2$ 
where $M,N=0,1,2,3,4$ are 5D spacetime indices, 
$\mu,\nu=0,1,2,3$ are 4D spacetime indices, 
$y=x_{M=4}$ represents the fifth coordinate and 
$\sigma(y)$ is an arbitrary function of $y$ 
(up to the following restriction). 
Because the fifth direction is compactified on $S^1/Z_2$, 
any field $f(x,y)$ (including gravity fields and then the 
above function $\sigma(y)$) satisfies $f(x,y+L)=f(x,y)$ 
and $f(x,-y)=f(x,y)$ (for $Z_2$ even fields) 
or $f(x,-y)=-f(x,y)$ (for $Z_2$ odd fields) 
where $L$ is the length of orbifold segment, and then 
there are two fixed points at $y=0$ and $y=L$. 

The supersymmetry in 5D has eight supercharges. 
For our purpose, relevant 5D supermultiplets 
are vector multiplets 
${\bm V}^I=\{ V^I, \Sigma^I \}$ 
with $I=1,2,\ldots,n_V$ 
and hypermultiplets 
${\bm H}_\alpha=\{ {\cal H}_\alpha, {\cal H}^C_\alpha \}$ 
with $\alpha=1,2,\ldots,n_H+n_C$, where 
$V^I$, $\Sigma^I$, ${\cal H}_\alpha$ and ${\cal H}^C_\alpha$ 
represent vector multiplets and three chiral multiplets, 
respectively, under the 4D supersymmetry preserved 
after the orbifolding which has four supercharges. 
We introduce multiple $Z_2$-odd vector multiplets 
${\bm V}^{I'}$ with $I'=1,2,\ldots,n_V'$ in which 
the zero-modes of $Z_2$-even chiral multiplets
$\Sigma^{I'}$ become moduli chiral multiplets $T^{I'}$ 
in 4D, and a linear combination\footnote{
In the case $n_V'=1$, the single modulus $T^{I'=1}$ 
corresponds to a so-called radion (chiral multiplet) 
satisfying $\langle {\rm Re}\,T^{I'=1} \rangle=L/\pi$, 
while the radion for $n_V'>1$ is a linear combination 
of $T^{I'}$s determined by the norm function 
(\ref{eq:nf}).} 
of these moduli becomes an inflaton 
field in the moduli inflation scenario. 
On the other hand, hypermultiplets ${\bm H}_\alpha$ are 
introduced in order to generate a suitable moduli 
potential for the inflation at the early universe 
and for a moduli stabilization at the present universe. 

The 5D off-shell (conformal) supergravity action 
for vector and hypermultiplets is completely fixed 
by identifying the numbers of multiplets, 
$n_V \ge 1$ and $n_H \ge n_C \ge 1$ where $n_C$ 
is the number of compensator hypermultiplets, 
and then determining a cubic polynomial of vector 
multiplets 
\begin{eqnarray}
{\cal N}(M) &=& 
\sum_{I,J,K=1}^{n_V}C_{I,J,K}M^IM^JM^K, 
\label{eq:nf}
\end{eqnarray}
with real coefficients $C_{I,J,K}$ 
for $I,J,K=1,2,\ldots,n_V$. 
The manifold of vector multiplets is called 
very special manifold governed by the norm function 
${\cal N}({\bm V})$, 
while that of hypermultiplets is dependent to $n_C$ 
(See Ref.~\cite{Zucker:1999ej,Kugo:2000af} and 
references therein). 
In this paper we choose $n_C=1$ for simplicity. 
The hypermultiplet ${\bm H}_\alpha$ can be a non-trivial 
representation of gauge symmetries in the 5D action 
whose gauge fields are identified with the vector 
fields $A_M^I$ in vector multiplets ${\bm V}^I$. 

In this paper we identify the $Z_2$-odd vector 
fields $A_M^{I'}$ in ${\bm V}^{I'}$ as gauge fields 
of $U(1)_{I'}$ symmetries for simplicity, 
and assign $U(1)_{I'}$ charges $c_{I'}^{(\alpha)}$ 
to the hypermultiplets ${\bm H_\alpha}$. 
We introduce the same number of 
(stabilizer) hypermultiplets $n_H=n_V'$ as that of 
$Z_2$-odd vector multiplets in order to stabilize 
the moduli $T^{I'}$ at a supersymmetric Minkowski 
minimum.\footnote{This moduli stabilization 
mechanism was proposed in Ref.~\cite{Maru:2003mq} 
to stabilize a single modulus, i.e., the radion 
for $n_V'=1$, which is extended here in this paper 
to the case with multiple moduli for $n_V'>1$.} 

So far we have set bulk configurations in the 
5D supergravity compactified on $S^1/Z_2$. 
In addition to these 5D data, 
K\"ahler and superpotential terms are allowed at 
the orbifold fixed points, where the supersymmetry 
is reduced to the one with four supercharges. 
For $Z_2$-even (stabilizer) chiral multiplets ${\cal H}_i$ 
($i=1,2,\ldots,n_H$) contained in the hypermultiplets 
${\bm H}_\alpha$, we consider in our scenario 
that the linear terms of ${\cal H}_i$, 
\begin{eqnarray}
{\cal W} &=& 
J_0^{(i)}\,{\cal H}_i\,\delta(y)
+J_L^{(i)}\,{\cal H}_i\,\delta(y-L), 
\label{eq:w5d}
\end{eqnarray}
where $J_{0,L}^{(i)}$ are constants, 
are dominant~\cite{Abe:2006eg} 
in the superpotential ${\cal W}$ induced at the 
fixed points $y=0$ and $y=L$, and the other terms 
are forbidden or negligible due to some symmetries 
or dynamics.\footnote{For $n_C=2$, a similar 
moduli stabilization potential was proposed~\cite{Abe:2007zv} 
in the framework of off-shell dimensional reduction.} 
Furthermore, we assume that the terms 
in the K\"ahler potential at the fixed points are 
also negligible compared with the bulk contributions, 
that can be a natural assumption if the radius $L/\pi$ 
of the compactified fifth dimension $y$ is larger 
enough than the inverse of the mass scales 
associated with these terms. 

Now the (off-shell) supergravity action in 5D 
spacetime is completely determined, and then 
employing the off-shell dimensional reduction~\cite{Abe:2006eg}, 
we can integrate it over the fifth dimension $y$ and 
find the following K\"ahler potential $K$ 
and the superpotential $W$, 
\begin{eqnarray}
K &=& -\ln {\cal N}({\rm Re}\,T)
+Z_{i,\bar{i}}({\rm Re}\,T)\,|H_i|^2, 
\nonumber \\
W &=& \left( J_0^{(i)} 
+e^{-c_{I'}^{(i)}T^{I'}}J_L^{(i)} \right) H_i, 
\label{eq:4dw}
\end{eqnarray}
in the 4D effective action, where 
\begin{eqnarray}
Z_{i,\bar{j}}({\rm Re}\,T) &=& 
\frac{1-e^{-2c_{I'}^{(i)}{\rm Re}\,T^{I'}}}{
c_{I'}^{(i)} {\rm Re}\,T^{I'}} \delta_{ij}, 
\label{eq:km}
\end{eqnarray}
is the K\"ahler metric of 4D zero-modes $H_{i}$ 
in a Kaluza-Klein expansion of $Z_2$-even components 
${\cal H}_i$ of hypermultiplet ${\bm H}_{\alpha=i}$. 
Here we remark that the exponential factors in $K$ 
and $W$ with the $U(1)_{I'}$ charges $c_{I'}^{(i)}$ 
in their exponents originate from the fact that the 
wavefunctions of zero-modes $H_i$ are localized 
exponentially in extra dimensions~\cite{Abe:2006eg}, 
which play important roles in this paper to realize 
a successful moduli inflation at the early universe 
as well as the moduli stabilization 
at the late time. 

We are ready to write down the effective 4D 
scalar potential $V$ for zero-modes $T^{I'}$ and $H_i$, 
where the former and the latter are called moduli and 
stabilizer fields (both chiral multiplets) respectively, 
using the standard formula of 4D supergravity 
(with four supercharges) as 
\begin{eqnarray}
V &=& e^K \left( 
K^{m,\bar{n}}\,D_mW\,D_{\bar{n}} \bar{W}
-3|W|^2 \right), 
\label{eq:4dsp}
\end{eqnarray}
where 
$D_m W=W_m+K_mW$, 
$W_m=\partial_m W$, 
$K_m=\partial_m K$, 
$m,n=\left\{ I', i \right\}$
and $K^{m,\bar{n}}$ 
is the inverse of K\"ahler metric 
$K_{m,\bar{n}}=\partial_m \partial_{\bar{n}}K$. 
The expectation values of moduli $T^{I'}$ 
and stabilizer fields $H_i$ at an extremum 
of the scalar potential~(\ref{eq:4dsp}) 
are found as~\cite{Maru:2003mq} 
\begin{eqnarray}
c_{I'}^{(i)} \langle T^{I'} \rangle 
&=& \ln \frac{J_L^{(i)}}{J_0^{(i)}}, \qquad 
\langle H_i \rangle \ = \ 0, 
\label{eq:movev}
\end{eqnarray}
which satisfy 
$\langle D_{I'}W \rangle 
= \langle D_iW \rangle 
= \langle W \rangle 
= 0$ 
and then 
$\langle V_{I'} \rangle 
= \langle V_i \rangle 
= \langle V \rangle 
= 0$
where 
$V_m=\partial_m V$. 

Without moduli mixings in the K\"ahler metric, 
$K_{I',\bar{J}'} = 0$ for $I' \ne J'$, 
fields are stabilized at a supersymmetric 
Minkowski minimum~(\ref{eq:movev}), 
where both the modulus $T^{I'}$ 
and the stabilizer field $H_i$ 
obtain a supersymmetric mass 
\begin{eqnarray}
m_{I'i}^2 &\simeq& 
\frac{e^{\langle K \rangle} 
\langle W_{I'i} \rangle^2}{
\langle K_{I',\bar{I}'} \rangle 
\langle K_{i,\bar{i}} \rangle}, 
\label{eq:thmass}
\end{eqnarray}
where 
$W_{mn}=\partial_m \partial_n W$ 
and then 
$\langle W_{I'i} \rangle 
=-c_{I'}^{(i)}e^{-c_{I'}^{(i)}
\langle T^{I'} \rangle}J_L^{(i)}$. 
It is remarkable that 
the mass square~(\ref{eq:thmass}) 
is exponentially suppressed with 
its exponent proportional to the 
$U(1)_{I'}$ charge $c_{I'}^{(i)}$ 
of the stabilizer field $H_i$, 
that is one of the consequences of 
wavefunction localization in the 
extra dimension as mentioned 
in Sec.~\ref{sec:intro}. 

Note that , for 
$\left| \langle W_{I'i} \rangle/\langle W_{J'j} \rangle \right| 
\sim {\cal O}(1)$ ($\forall I',J',i,j$),  this moduli stabilization 
mechanism does not work with a sizable 
moduli mixing in the K\"ahler metric, 
$K_{I',\bar{J}'} \ne 0$ for $I' \ne J'$, 
with which the above expectation 
values~(\ref{eq:movev}) correspond 
to a saddle point or a local maximum 
of the scalar potential. 
Therefore, the coefficients $C_{I',J',K'}$ 
in the norm function~(\ref{eq:nf}) 
are restricted\footnote{In the later 
concrete example of our model 
in Sec. \ref{sec:sinflation}, we will 
take the norm function~(\ref{eq:nfinf}) 
and~(\ref{eq:qpnv2}) for $n_V'=2$, 
that leads to such a diagonal metric.} 
to those yielding an almost diagonal 
moduli K\"ahler metric, 
$K_{I',\bar{J}'} \approx 0$ for $I' \ne J'$ 
at least at the minimum~(\ref{eq:movev}). 
On the other hand, for 
$\left| \langle W_{I'i} \rangle/\langle W_{J'j} \rangle \right| 
\ll 1$ ($\exists I',J',i,j$), 
it is possible that a sizable K\"ahler 
mixing does not spoil the stability of 
the vaccum (\ref{eq:movev}) depending 
on the hierarchy of $\langle W_{I'i} \rangle$. 
The large-field model proposed in Sec. 
\ref{sec:linflation} 
utilizes this fact. 

From a particle phenomenological point 
of view, we have to introduce a supersymmetry 
breaking sector, which in general affects 
the moduli stabilization. 
For the case presented in this paper, 
the shift of position of 
the minimum~(\ref{eq:movev}) 
is negligible if 
the supersymmetric mass~(\ref{eq:thmass}) 
is larger enough than the supersymmetry 
breaking scale. Even in this case, 
the height of the minimum will be affected by 
the supersymmetry breaking, and we just assume 
$\langle V \rangle \approx 0$ even after the 
breaking sector is incorporated. 
We will discuss the validity 
of this assumption in Sec. \ref{subsec:modpro}.

\section{A simple model for the small-field inflation}
\label{sec:sinflation}
In this section, we show that the moduli 
potential discussed so far allows a scenario 
of small-field moduli inflation that can realize the 
observed WMAP and Planck data~\cite{Ade:2013zuv}. 

We consider the case that 
one pair of modulus and stabilizer fields, 
e.g. $T^{I'=1}$ and $H_{i=1}$, 
is decoupled from and lighter enough 
than the other pairs 
$T^{I' \ne 1}$ and $H_{i \ne 1}$, 
that is, 
$m_{I'=1,i=1}^2 \ll m_{I' \ne 1,i \ne 1}^2$. 
Such a case can be naturally realized when 
$c_{I'=1}^{(i \ne 1)}=c_{I' \ne 1}^{(i=1)}=0$ 
in Eq.~(\ref{eq:4dw}) and 
$\left| c_{I' \ne 1}^{(i \ne 1)} \right|
<\left| c_{I'=1}^{(i=1)} \right|$ 
in Eq.~(\ref{eq:thmass}). 
In this case, below the heavier mass scale 
$m_{I' \ne 1,i \ne 1}$, 
all the moduli $T^{I' \ne 1}$ 
and stabilizer fields $H_{i \ne 1}$ 
except the lighter pair 
$T^{I'=1}$ and $H_{i=1}$ 
are strictly fixed at their supersymmetric 
minimum with no fluctuations around it, 
and they are replaced by their vacuum 
expectation values~(\ref{eq:movev})
in the low energy effective action. 
Then, the effective K\"ahler potential 
and superpotential for the lighter fields 
$T^1$ and $H_1$ are given by 
\begin{eqnarray}
K_{\rm eff}(T^1,H_1) 
&=& K(T^{I'},H_i)\,\Big|_0 
\ = \ -\ln {\cal N}({\rm Re}\,T)\,\Big|_0
+Z_{1,\bar{1}}
({\rm Re}\,T)\,\Big|_0 \,|H_1|^2, 
\nonumber \\
W_{\rm eff}(T^1,H_1) 
&=& W(T^{I'},H_i)\,\Big|_0 
\ = \ \left( J_0^{(1)} 
+e^{-c_1^{(1)}T^1}J_L^{(1)} \right) H_1, 
\label{eq:weff}
\end{eqnarray}
respectively, where 
$f(T^{I'},H_i)\,\Big|_0 \equiv f(T^{I'},H_i)
\Big|_{\scriptsize \begin{array}{ll}
T^{I' \ne 1}=\langle T^{I' \ne 1} \rangle \\ 
H_{i \ne 1}=\langle H_{i \ne 1} \rangle
\end{array}}$ 
for an arbitrary function 
$f(T^{I'},H_i)$, and then 
\begin{eqnarray}
Z_{1,\bar{1}}({\rm Re}\,T)\,\Big|_0 &=& 
\frac{1-e^{-2c_1^{(1)}{\rm Re}\,T^1}}{
c_1^{(1)} {\rm Re}\,T^1}. 
\nonumber
\end{eqnarray}
The effective potential for the light fields, 
\begin{eqnarray}
V_{\rm eff}(T^1,H_1) &=& e^{K_{\rm eff}} 
\left( (K_{\rm eff})^{m,\bar{n}}\,
D_mW_{\rm eff}\,D_{\bar{n}} \bar{W}_{\rm eff}
-3|W_{\rm eff}|^2 \right), 
\label{eq:4dspeff}
\end{eqnarray}
is obtained by using the effective K\"ahler potential 
and superpotential~(\ref{eq:weff}), 
where $m,n=\{I',i\}$ with $I'=1$ and $i=1$. 

\subsection{The inflaton potential}
\label{ssec:potential}
\begin{figure}[t]
\centering \leavevmode
\includegraphics[width=0.6\linewidth]{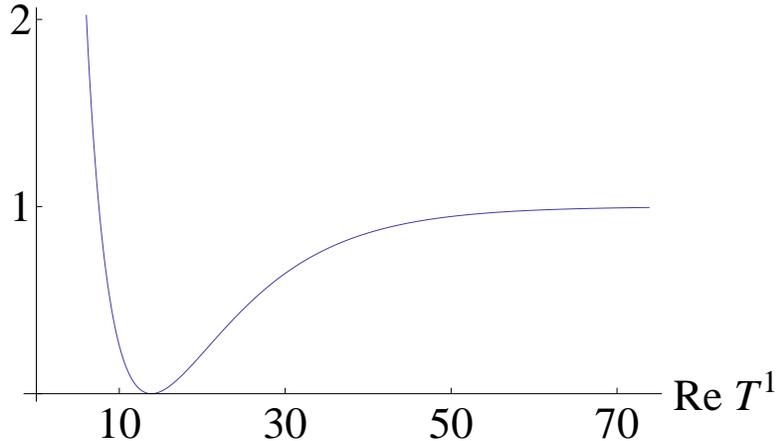}
\caption{Scalar potential 
$V_{\rm eff}(T^1,H_1)/V_\infty$ 
on the ${\rm Im}\,T^1=H_1=0$ hypersurface}
\label{fig:ip}
\end{figure}

From Eq.\,(\ref{eq:4dspeff}), we find the effective 
potential for the modulus $T^1$ (whose real part 
will be identified as the inflaton field later) 
on the $H_1=0$ hypersurface, 
\begin{eqnarray}
V_{\rm eff}(T^1,H_1=0) 
&=& e^{K_{\rm eff}}
(K_{\rm eff})^{i=1,\bar{i}=\bar{1}} 
\left| (W_{\rm eff})_{i=1} \right|^2 
\,\Big|_{H_1=0}
\nonumber \\ &=& 
\frac{c_1^{(1)} {\rm Re}\,T^1}{
{\cal N}({\rm Re}\,T)\,\Big|_0}
\times 
\frac{\left| J_0^{(1)} \right|^2 
\left| 1+\frac{J_L^{(1)}}{J_0^{(1)}} 
e^{-c_1^{(1)}T^1} \right|^2}{
1-e^{-2c_1^{(1)}{\rm Re}\,T^1}}, 
\label{eq:veff}
\end{eqnarray}
where 
$(K_{\rm eff})^{i,\bar{i}}\Big|_{H_1=0}
=1/Z_{i,\bar{i}}({\rm Re}\,T)\,\Big|_0$. 
In the case 
\begin{eqnarray}
{\cal N}({\rm Re}\,T)\Big|_0 
&=& {\cal P}_0\,{\rm Re}\,T^1, 
\label{eq:cdnf}
\end{eqnarray}
where ${\cal P}_0$ does not depend on $T^1$, 
the first factor in Eq.~(\ref{eq:veff}) 
is independent to $T^1$, and then we find 
\begin{eqnarray}
\lim_{{\rm Re}\,T^1 \to 0} 
\left| V_{\rm eff}(T^1,H_1=0) \right| 
&=& \infty, \nonumber \\
\lim_{{\rm Re}\,T^1 \to \infty} 
V_{\rm eff}(T^1,H_1=0) &=& 
c_1^{(1)} {\cal P}_0^{-1} 
\left| J_0^{(1)} \right|^2 
\ \equiv \ V_\infty, 
\label{eq:limveff}
\end{eqnarray}
for $J_L^{(1)}/J_0^{(1)} \ne -1$ and $c_1^{(1)}>0$. 
Fig.~\ref{fig:ip} shows 
the ${\rm Re}\,T^1$-dependence of 
$V_{\rm eff}(T^1,H_1)/V_\infty$ 
on the ${\rm Im}\,T^1=H_1=0$ hypersurface, 
where the parameters are chosen as 
\begin{eqnarray}
c_1^{(1)} &=& 1/10, \qquad 
J_L^{(1)}/J_0^{(1)} \ = \ -3.9, \qquad 
J_0^{(1)} \ = \ 10^{-4}, 
\label{eq:para}
\end{eqnarray} 
in the Planck scale unit $M_{\rm Pl}=1$. 
In Fig.~\ref{fig:ip}, we recognize the above 
feature~(\ref{eq:limveff}) and expect that ${\rm Re}\,T^1$ 
can play a role of inflaton field, starting from its large 
positive value on the flat region of the potential and slowly 
rolling down to the minimum\footnote{We comment that this 
shape of potential is essentially the same as the one in 
Starobinski model~\cite{Starobinsky:1980te}, 
but the origin of the potential is quite different.} 
given by Eq.~(\ref{eq:movev}) for $i=1$. 
Furthermore, we find that the overshooting to negative region 
${\rm Re}\,T^1<0$ is prohibited, which is also understood 
from Eq~(\ref{eq:limveff}). 

Before analyzing the inflation dynamics, we should 
recall the fact that the flatness of the potential 
in the large ${\rm Re}\,T^1$ region is guaranteed 
by the assumption~(\ref{eq:cdnf}). 
The most general form of norm function~(\ref{eq:nf}) 
satisfying the condition~(\ref{eq:cdnf}) is found as 
\begin{eqnarray}
{\cal N}(M) &=& {\cal P}(M)\,M^1+\cdots, 
\label{eq:nfinf}
\end{eqnarray}
where 
\begin{eqnarray}
{\cal P}(M) &=& 
\sum_{J', K' \ne 1}^{n_V'} C_{1,J',K'}M^{J'}M^{K'}, 
\label{eq:qp}
\end{eqnarray}
is a quadratic polynomial of fields $M^{I' \ne 1}$
in $Z_2$-odd vector multiplets ${\bm V}^{I' \ne 1}$ 
other than ${\bm V}^{I'=1}$, 
and the ellipsis stands for terms including fields 
$M^{I''}$ in $Z_2$-even vector multiplets 
${\bm V}^{I''}=\{V^{I''},\Sigma^{I''}\}$ 
with $I''=n_V'+1,n_V'+2,\ldots,n_V$ 
whose components $\Sigma^{I''}$ are $Z_2$-odd 
chiral multiplets which do not carry any moduli. 
The coefficient ${\cal P}_0$ of ${\rm Re}\,T^1$ 
in Eq.~(\ref{eq:cdnf}) is given by 
${\cal P}_0 = {\cal P}({\rm Re}\,T)\Big|_0$, 
which is a field independent constant 
by definition~(\ref{eq:qp}). 

Therefore, we find that the interesting flat region 
is realized in a moduli stabilization potential 
generated by a simple superpotential~(\ref{eq:w5d}) 
as a consequence of the peculiar form of norm 
function~(\ref{eq:nfinf}) in 5D supergravity. 
Note especially that, the condition~(\ref{eq:nfinf}) 
cannot be satisfied for $n_V'=1$, i.e., the single 
modulus case where only the radion exists, because 
the norm function ${\cal N}(M)$ is a cubic polynomial. 
For $n_V'=2$, the quadratic polynomial ${\cal P}(M)$ 
is uniquely determined as 
\begin{eqnarray}
{\cal P}(M) &=& C_{1,2,2}\,(M^2)^2. 
\label{eq:qpnv2}
\end{eqnarray}

Finally we remark that, 
although the norm function coefficients $C_{I,J,K}$ 
are free parameters in 5D supergravity, these are 
closely related to the structure of the internal 
manifold, if it is the 5D effective theory of a 
more fundamental theory defined in more than five 
dimensional spacetime with extra dimensions 
compactified on some manifold.\footnote{
One of such examples is the 5D effective 
theory of heterotic M-theory, 
where the norm function coefficients correspond 
to the intersection numbers of internal Calabi-Yau 
three-fold~\cite{Lukas:1998yy}.} 
In such a situation, the cosmological (as well as 
phenomenological) 
features of 5D supergravity are governed by the 
internal manifold behind it. 
  
\subsection{The inflation dynamics}
\label{ssec:inflation}
Based on the previous arguments, we identify one 
of the moduli fields, the real part of the lightest 
modulus, ${\rm Re}\,T^1$, as the inflaton field. 
Although the inflation mechanism proposed in this 
paper is applicable to any number $n_V' \ge 2$ of 
$Z_2$-odd vector multiplets ${\bm V}^{I'}$, 
in the following, we choose the minimal number 
$n_V'=2$ just for simplicity and concreteness. 
Then the norm function~(\ref{eq:nfinf}) 
is uniquely determined by the quadratic 
monomial~(\ref{eq:qpnv2}), where we set 
$C_{1,2,2}=1$ without loss of generality 
which determines the normalization 
of the field $M^2$. 

By assuming that oscillations of 
the other light fields 
${\rm Im}\,T^1$, 
${\rm Re}\,H_1$ and 
${\rm Im}\,H_1$ 
than ${\rm Re}\,T^1$ 
around their expectation values 
$\langle {\rm Im}\,T^1 \rangle 
=\langle {\rm Re}\,H_1 \rangle 
=\langle {\rm Im}\,H_1 \rangle=0$ 
are negligible during and after the inflation 
(which will be confirmed in Sec. \ref{subsec:modpro}), 
we solve the equation of motion for the single field 
$\sigma \equiv {\rm Re}\,T^1$, 
\begin{eqnarray}
\ddot\sigma +3H \dot\sigma 
+\Gamma^\sigma_{\ \sigma \sigma} \dot\sigma^2 
+g^{\sigma \sigma} 
\frac{\partial V_{\rm eff}}{\partial \sigma}
&=& 0, 
\label{eq:eominf}
\end{eqnarray}
where the dot denotes the derivative $\frac{d}{dt}$ 
with respect to a cosmic time $t$, $V_{\rm eff}$ 
is the effective potential~(\ref{eq:4dspeff}), 
$g_{\sigma \sigma}=2(K_{\rm eff})_{I'=1,J'=1}$, 
$g^{\sigma \sigma}=g_{\sigma \sigma}^{-1}$ and 
$\Gamma^\sigma_{\ \sigma \sigma}$ is the 
Christoffel symbol constructed by the metric 
$g_{\sigma \sigma}$, all on the 
${\rm Im}\,T^1={\rm Re}\,H_1={\rm Im}\,H_1=0$ 
hypersurface of the field space. 
The Hubble parameter $H$ is given as 
$H^2=\left( \frac{\dot{a}}{a} \right)^2
=\frac{1}{6} g_{\sigma \sigma} \dot\sigma^2 
+\frac{V_{\rm eff}}{3}$ where 
$a$ is the scale factor of 4D spacetime, 
in which the 4D effective theory of 
5D supergravity is defined. 

Eq.~(\ref{eq:eominf}) is rewritten as 
\begin{eqnarray}
\sigma'' 
&=& -\left( 1-
\frac{g_{\sigma \sigma}(\sigma')^2}{6} \right) 
\left( 3\sigma'+6\sigma^2 
\frac{V_{\rm eff}'}{V_{\rm eff}} \right) 
+\frac{(\sigma')^2}{\sigma}, 
\label{eq:eominfe}
\end{eqnarray}
where the prime denotes the derivative 
$\frac{d}{dN}=H^{-1}\frac{d}{dt}$ 
with respect to the number $N \equiv \ln a(t)$ 
of e-foldings, and we have used 
$\Gamma^\sigma_{\ \sigma \sigma}=-1/\sigma$. 

In the following analysis, 
the numerical values of parameters 
in the Planck scale unit $M_{\rm Pl}=1$ 
are chosen as 
\begin{eqnarray}
c_2^{(2)} &=& 1/20, \qquad 
J_L^{(2)}/J_0^{(2)} \ = \ -9, \qquad 
J_0^{(2)} \ = \ 10^{-1}, 
\label{eq:parah}
\end{eqnarray} 
for the heavy fields $T^2$ and $H^2$ 
as well as those~(\ref{eq:para}) 
for the light fields $T^1$ and $H^1$. 
With these parameters, 
the vacuum expectation values of fields 
are given by Eq.~(\ref{eq:movev}), 
and their numerical values are found as 
\begin{eqnarray}
\langle T^1 \rangle \ \simeq \ 13.6, \qquad 
\langle T^2 \rangle \ \simeq \ 43.9, \qquad 
\langle H_1 \rangle \ = \ 
\langle H_2 \rangle \ = \ 0, 
\nonumber
\end{eqnarray}
that determine 
\begin{eqnarray}
{\cal P}_0 
&=& {\cal P}({\rm Re}\,T)\Big|_0
\ = \ \langle {\rm Re}\,T^2 \rangle^2. 
\nonumber
\end{eqnarray}

At this supersymmetric Minkowski minimum, the 
supersymmetric mass squares~(\ref{eq:thmass}) 
of light ($I'=1$, $i=1$) 
and heavy ($I'=2$, $i=2$) fields 
are estimated respectively as 
\begin{eqnarray}
m^2_{I'=1,i=1} 
& \simeq & \left( 4.9 \times 10^{12} \ 
{\rm GeV} \right)^2, \qquad 
m^2_{I'=2,i=2} 
\ \simeq \ \left( 6.9 \times 10^{15} \ 
{\rm GeV} \right)^2, 
\nonumber
\end{eqnarray}
while the inflation scale in our model is characterized 
by the Hubble scale
\begin{eqnarray}
H_{\rm inf} &\equiv &\left( V_\infty /3M_{\rm Pl}^2\right)^{1/2} 
\ \simeq \ 1.0 \times 10^{12} \ {\rm GeV}, 
\label{eq:sfhinf}
\end{eqnarray}
with $V_\infty$ given in Eq.~(\ref{eq:limveff}), 
for $M_{\rm Pl}=2.4 \times 10^{18}$ GeV. 
Because all of these scales $m_{I'=1,i=1}$, 
$m_{I'=2,i=2}$ and $H_{\rm inf}$ are 
below the compactification scale 
\begin{eqnarray}
M_C &\equiv& \frac{\pi}{L} 
\ \simeq \ \frac{\pi M_{\rm Pl}}{\langle 
{\cal N}({\rm Re}\,T) \rangle^{1/2}} 
\ \simeq \ 4.7 \times 10^{16} \ {\rm GeV}, 
\nonumber
\end{eqnarray}
we find the parameters chosen here ensure 
the validity of 4D effective-theory 
description during and after the inflation. 
It is also confirmed that the heavy fields 
$T^2$ and $H_2$ are decoupled from the 
inflation dynamics due to $m_{I'=1,i=1} 
\sim H_{\rm inf} \ll m_{I'=2,i=2}$, 
and their oscillations can be neglected. 

Now we consider a possibility of slow roll 
inflation starting from a large value of 
inflaton field $\sigma$ in the flat region of 
the potential down to its VEV~(\ref{eq:movev}) 
at the minimum. To estimate the observed quantities, 
we define the generalized slow roll parameters 
for the 
scalars having non-canonical kinetic term 
\cite{Burgess:2004kv}, 
 \begin{align}
\epsilon &\equiv \frac{M_{Pl}^2}{2} 
\frac{\partial_\sigma V_{\rm eff} 
g^{\sigma\sigma} \partial_\sigma V_{\rm eff}}{V_{\rm eff}^2}
\sim (2c_1^{(1)}\sigma)^2  \left(\frac{J_L^{(1)}}{J_0^{(1)}}e^{-c_1^{(1)}\sigma}\right)^2,
\nonumber\\
\eta &\equiv \frac{\nabla^\sigma\nabla_\sigma V_{\rm eff}}{V_{\rm eff}} 
= \frac{g^{\sigma\sigma}\partial_\sigma^2 V_{\rm eff} 
-g^{\sigma\sigma}\Gamma^\sigma_{\sigma\sigma} \partial_\sigma V_{\rm eff}}
{V_{\rm eff}}
\sim -(2c_1^{(1)} \sigma)^2  \frac{J_L^{(1)}}{J_0^{(1)}}e^{-c_1^{(1)}\sigma},
\end{align}
where $\nabla_\sigma$ is the covariant derivative 
for the field $\sigma$. The observables such as the 
power spectrum of scalar curvature perturbation, 
its spectral index and the tensor-to-scalar ratio 
are written in terms of these slow-roll parameters 
as
\begin{align}
P_\xi (k)&=\frac{1}{24\pi^2} \frac{V}{\epsilon M_{\rm Pl}^4}, \nonumber\\
n_s &= 1+ \frac{d\,{\rm ln} P_{\xi}(k)}{d\,{\rm ln}\,k} 
\simeq 1-6\,\epsilon +2\,\eta, \nonumber\\
r &=16\,\epsilon,
\label{eq:obss}
\end{align} 
 
\begin{figure}[h]
\begin{minipage}{0.5\hsize}
\begin{center}
\includegraphics[width=0.8\linewidth]{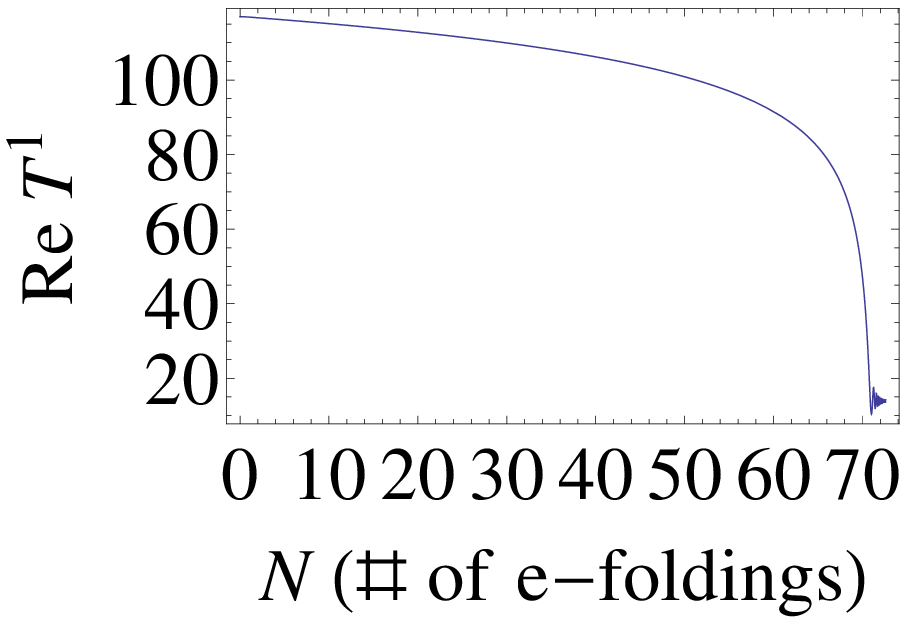}
\end{center}
\end{minipage}
\begin{minipage}{0.5\hsize}
\begin{center}
\includegraphics[width=0.8\linewidth]{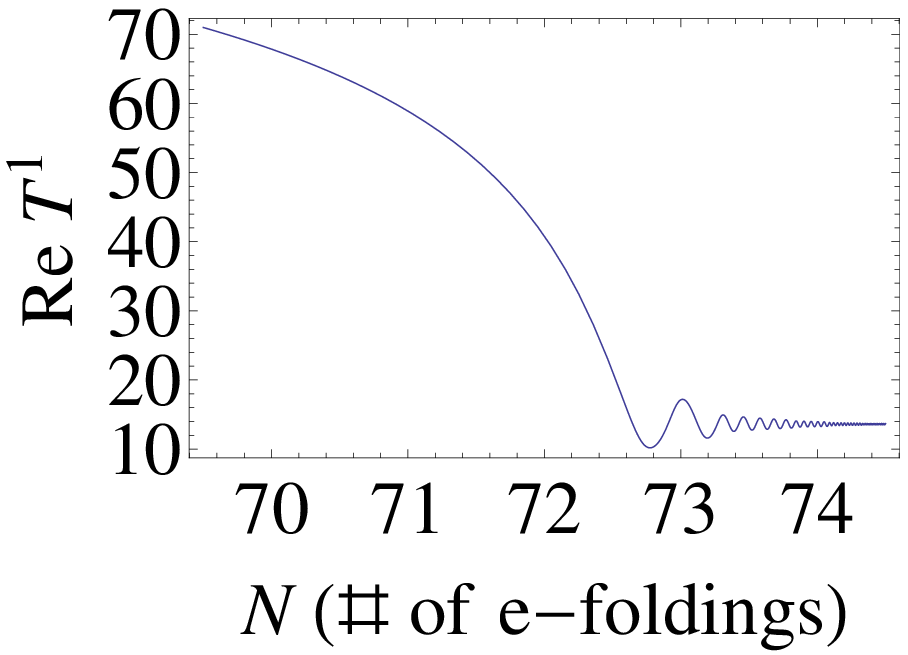}
\end{center}
\end{minipage}
\caption{The behavior of inflaton field 
$\sigma={\rm Re}\,T^1$ as a function 
of the e-folding number $N$.}
\label{fig:io}
\end{figure}

We numerically solve Eq.~(\ref{eq:eominfe}) 
with the initial conditions $\sigma=117$ and 
$\sigma'=0$ at $N=0$. Fig.~\ref{fig:io} shows 
the evolution of $\sigma$ as a function of $N$. 
In this figure, we find the inflation ends 
at about $N_{\rm end} \simeq 70.7$ where 
the slow-roll condition 
is violated (max $\{\epsilon, \eta\}\,=\,1$) 
and then 
the oscillation of inflaton starts.

First, we denote the field value 
$\sigma=\sigma_\ast$ 
corresponding to the pivot scale 
$k_0=0.05$ [Mpc$^{-1}$] 
(at which the horizon exits) and 
the scalar potential 
$V_\ast^{1/4} 
\equiv V^{1/4}(\sigma_\ast)$ 
at the pivot scale and 
$V_{\rm end}^{1/4} 
\equiv V^{1/4}(\sigma_{\rm end})$ 
at the end of inflation. In terms of 
them, the e-folding number after 
the pivot scale is given by~\cite{Liddle:1993fq},
\begin{eqnarray}
N_{\rm e} &\equiv& N_{\rm end}-N_\ast 
\simeq \, 62+\ln 
\frac{V_\ast^{1/4}}{10^{16}\ {\rm GeV}} 
+\ln \frac{V_\ast^{1/4}}{V_{\rm end}^{1/4}} 
-\frac{1}{3} \ln 
\frac{V_{\rm end}^{1/4}}{\rho_R^{1/4}} 
\ \simeq \ 56, 
\label{eq:nend}
\end{eqnarray}
where $V_\ast^{1/4} \simeq V_{\rm end}^{1/4} 
\simeq 2 \times 10^{15}$ GeV 
and $\rho_R$ is the energy density 
by which the universe is thermalized 
with the reheating temperature 
$T_R \simeq 1.05 \times 10^9$ GeV 
whose numerical value will be 
determined later in Sec.~\ref{subsec:rehs}. 
Note that the energy of inflaton is assumed, 
in Eq.~(\ref{eq:nend}), to be instantaneously 
converted into radiation. On the other hand, 
the same number $N_{\rm e}$ is estimated 
based on a slow-roll approximation, 
\begin{eqnarray}
N_{\rm e} \ = \ -\int_{t_{\rm end}}^{t_\ast} 
d\tilde{t}\, H(\tilde{t})
\ \simeq \ \frac{1}{M_{\rm Pl}^2} 
\int_{\sigma_{\rm end}}^{\sigma_\ast} d\sigma 
\frac{V_{\rm eff}}{g^{\sigma \sigma}V'_{\rm eff}}, 
\label{eq:nast}
\end{eqnarray}
and then we find the numerical value 
\begin{eqnarray}
\sigma_\ast &\simeq& 114, 
\label{eq:sast}
\end{eqnarray}
is determined by equaling the Eq. (\ref{eq:nend}) 
and Eq. (\ref{eq:nast}).

Second, we check whether the WMAP and Planck 
normalization on the power spectrum of scalar 
curvature perturbation, $P_\xi(k_0)
=2.196_{-0.060}^{+0.051} \times 10^{-9}$
~\cite{Ade:2013zuv}, 
can be realized or not. 
The slow-roll parameters $\epsilon$ and $\eta$ 
are obtained at the pivot scale $k_0$ by using the 
numerical value (\ref{eq:sast}),
\begin{align}
\epsilon &\sim (2c_1^{(1)}\sigma)^2  
\left(\frac{J_L^{(1)}}{J_0^{(1)}}e^{-c_1^{(1)}
\sigma}\right)^2\Bigl|_{\sigma =\sigma_\ast} \simeq 
{\cal O}(10^{-6}), 
\nonumber\\
\eta &\sim -(2c_1^{(1)} \sigma)^2  
\frac{J_L^{(1)}}{J_0^{(1)}}e^{-c_1^{(1)}\sigma}\Bigl|_{\sigma =\sigma_\ast} 
\simeq {\cal O}(-0.02),
\label{eq:slowpara}
\end{align}
which yield $P_\xi(k_0) \sim 2 \times 10^{-9}$ of 
the correct order of  the observed value. 
Inversely speaking, the parameters $J_0^{(1)}$ and 
$J_L^{(1)}$ are determined in Eq. (\ref{eq:para}) 
in such a way that the resultant $P_\xi(k_0)$ resides 
in the observed region. 
Also, the spectral index of the scalar curvature 
perturbation, $n_s \,=0.9603\pm 0.0073$
~\cite{Ade:2013zuv}, at 
the pivot scale is observed by the WMAP and 
Planck collaborations. In our model, we can 
realize the correct value of the spectral index, 
$n_s \,\simeq\,0.96$ by using Eq. (\ref{eq:obss}) 
and Eq. (\ref{eq:slowpara}). 
It implies that the $\eta$ problem is avoided by 
the exponential factor and the large value of the 
inflaton field, because the shift symmetry of 
$\sigma$ is violated by its own superpotential 
(\ref{eq:weff}).

\begin{figure}[h]
\centering \leavevmode
\includegraphics[width=0.5\linewidth]{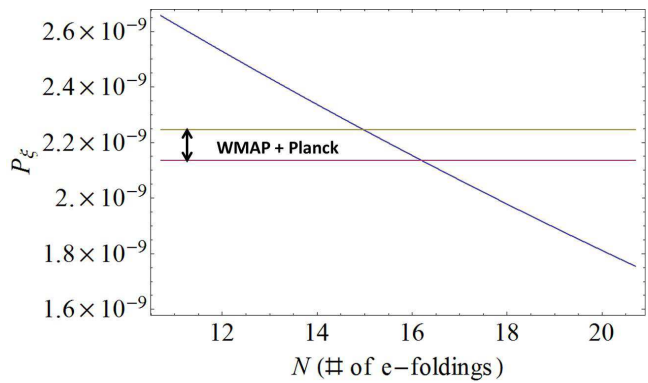}
\caption{The power spectrum of scalar curvature 
perturbation between 60 and 50 
e-foldings before the end of inflation.}
\label{fig:ps}
\end{figure}

\begin{figure}
\centering \leavevmode
\includegraphics[width=0.5\linewidth]{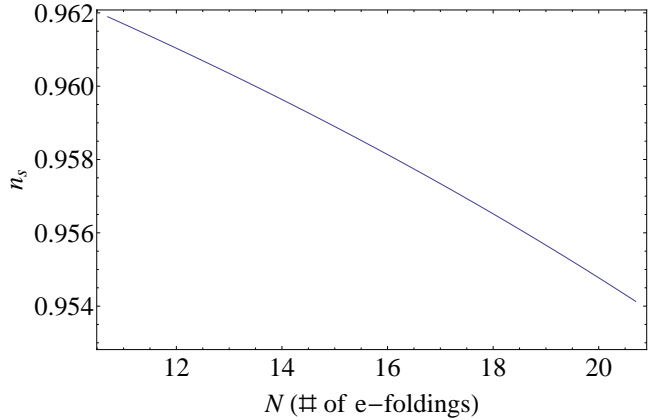}
\caption{The spectral index of scalar curvature 
perturbation between 60 and 50 
e-foldings before the end of inflation.}
\label{fig:si}
\end{figure}

We summarize the results of inflation dynamics 
in Figs. \ref{fig:ps} and \ref{fig:si}. 
From these figures drawn with the sample values 
of parameters (\ref{eq:para}) and (\ref{eq:parah}), 
we extract the numerical values of observables as 
\begin{align}
N_e= 55.6 \;\text{e-folds},\qquad
P_\xi = 2.23\times 10^{-9},\qquad 
n_s =0.959, \qquad 
r=1.6\times 10^{-5}.
\label{eq:res}
\end{align}
So the simple model analyzed so far with 
Re\,$T^1$ playing the role of inflation 
is consistent with the WMAP and 
Planck data~\cite{Ade:2013zuv}. 
Note that this inflation mechanism 
is categorized as the small-field model 
of inflation due to the tiny 
slow-roll parameter $\epsilon$ and the 
field variable of the inflaton spends
\begin{align}
\Delta\hat{\sigma} \equiv \hat{\sigma}_\ast 
-\hat{\sigma}_{\rm end} 
\simeq 0.77 M_{\rm Pl},\;\;\; \hat{\sigma}=\frac{1}{2} {\rm log}\,\sigma,
\label{eq:infcans}
\end{align}
where we canonically normalize the field 
$\sigma ={\rm Re}\,T^1$. 
This small-field inflation leads to the small 
tensor-to-scalar ratio as can be seen in 
Eq. (\ref{eq:res}). 
Although the results here are completely 
consistent with WMAP and the current 
Planck data, the tiny tensor-to-scalar ratio 
$r\sim 10^{-5}$ in Eq. (\ref{eq:res}) 
contradicts with the most recent data from 
the BICPE2 collaborations~\cite{Ade:2014xna}. 
We discuss how to realize a successful large-field 
inflation in Sec. \ref{sec:linflation}, which is one 
of the few candidates to generate a sizable 
tensor-to-scalar ratio within the framework 
of single-field slow-roll inflation.

\vspace{4cm}
\subsection{Reheating temperature}
\label{subsec:rehs}
Before analyzing the large-field inflation, we comment on 
the process to reheat the universe. 
After the inflation, the energy of the inflaton is reduced 
via inflaton decay into particles in the supersymmetric 
standard model, although it depends on the concrete model. 
Here we roughly estimate the reheating temperature via 
the decay from the 
inflaton into gauge boson pairs due to the dimensional 
counting. 

If the particles in the supersymmetric standard 
model have the $U(1)_{I'=1}$ charge for the 
vector multiplet $V^{I'=1}$ carrying inflaton, there are 
terms like $Z_{Q,\bar{Q}}({\rm Re}\,T)|Q|^2$ in the 
K\"ahler potential with $Q$ being the matter chiral 
multiplet originated from the hypermultiplet 
${\bm H}_\alpha$ and $Z_{Q,\bar{Q}}({\rm Re}\,T)$ 
is the K\"ahler metric of $Q$ given by Eq. (\ref{eq:km}) 
where $c_{I'}^{(i)}$ is replaced by the $U(1)_{I'}$ charge of 
$Q$. 
Although these couplings may enhance the inflaton  
decay width into $Q$, we will not consider them in this 
paper for simplicity just assuming the vanishing 
$U(1)_{I'=1}$ charges for matter fields. 
\footnote{If the $U(1)_{I'=1}$ charge of $Q$ is of 
${\cal O}(1)$, the decay width into $Q$ is almost 
the same order as those into the gauge boson 
pairs, $\Gamma (\hat{\sigma} \rightarrow g^{(a)}+g^{(a)})$.} 
The couplings between moduli and gauge fields 
are conducted by the gauge kinetic function $f_a(T)$, 
where $a=1,2,3$ represents the gauge groups in the 
minimal supersymmetric standard model (MSSM), 
$U(1)_Y$, $SU(2)_L$, $SU(3)_c$ respectively. 
The relevant terms in the Lagrangian are 
\begin{align}
{\cal L}&\supset -\cfrac{1}{4} {\rm Re}\,f_a 
F^a_{\mu\nu}F^{a\mu\nu} \nonumber\\
&= -\cfrac{1}{4}\langle {\rm Re}\,f_a \rangle 
F^a_{\mu\nu}F^{a\mu\nu} 
-\cfrac{1}{4}\left\langle\frac{\partial\,{\rm Re}\,f_a}
{\partial \hat{\sigma}}\right\rangle 
\delta \hat{\sigma} F^a_{\mu\nu}F^{a\mu\nu},
\end{align}
where $f_a =\sum_{I'=1}^2 \xi_a^{I'} T^{I'}$ and 
$\xi_a^{I'} \equiv C_{I^{'},\,J^{''}=a,\,K^{''}=a}$. 
Then the total decay width of the inflaton $\hat{\sigma}$ is 
approximated as 
\begin{align}
\Gamma\,\simeq\,\sum_{a=1}^3
\Gamma (\hat{\sigma} \rightarrow g^{(a)}+g^{(a)})&= 
\sum_{a=1}^3\cfrac{N_G^a}{128\pi} \left\langle\cfrac{\xi_{a}^1}{\sqrt{(K_{\rm eff})_{T^1T^1}}
{\rm Re}\,f_a}\right\rangle^2
\cfrac{m_{\hat{\sigma}}^3}{M_{Pl}^2}, \nonumber\\
&\simeq \,2.32 \,{\rm GeV},
\end{align}
where $\sum_{a=1}^3\,N_G^a\,=\,12$ is the number of the 
gauge boson in the MSSM and $\hat{\sigma}$ is the canonically 
normalized inflaton field (\ref{eq:infcans}). 
We choose the $\xi_1^1=\xi_2^1=\xi_3^1=0.27$, 
otherwise zero to realize the gauge coupling unification 
at the grand unification scale ($\simeq \,2.0\times 10^{16} {\rm GeV}$), 
\begin{align}
\text{Re}\,f_a(\langle T \rangle) =
\left(\frac{1}{g_a}\right)^2 \simeq \,3.73. 
\end{align}

Then the reheating temperature is roughly estimated 
by equaling the expansion rate of the universe and the total decay width, 
\begin{align}
\Gamma &\simeq \,H (T_R), \nonumber\\
\Leftrightarrow T_{R} &= \left( \cfrac{\pi^2 g_\ast}{90}\right)^{-1/4} 
\sqrt{\Gamma \,M_{\rm Pl}} \simeq  1.05\times 10^9\,{\rm GeV},
\end{align}
where we use $g_\ast=915/4$ which is the 
effective degrees of freedom of the radiation at 
the reheating in the MSSM.
We restrict ourselves to the standard situation 
that the coherent oscillation of inflaton field 
dominates the energy density of the universe 
after the inflation. The inflaton releases the 
entropy and reheats the universe when it decay. 
It is then assumed that the other field does 
not dominate the energy density of the universe 
which is verified in Sec. \ref{subsec:modpro}. 

Finally in this section, we mention about the 
one-loop correction to the moduli K\"ahler 
potential. The modified K\"ahler potential in 
the large volume limit is found as~\cite{Sakamura:2013wia}, 
\begin{align}
K =-\ln\,{\cal N} 
+{\cal O}\left( \frac{1}{32\pi^2{\cal N}}\right) +\cdots,
\label{eq:oneloop}
\end{align}
where the leading contribution will depend on 
the number of the charged fields under the 
$Z_2$-odd vector multiplets $V^{I'}$. 
Even if there are such contributions in the scalar potential, 
our estimation in the previous section is not 
changed due to the supersymmetry condition 
(\ref{eq:movev}) at the vacuum.
Since Re\,$T^1$ rolls the potential from the 
large field value, one-loop effect does not 
affect the inflation mechanism, which is also 
confirmed by the numerical analysis.

\section{A simple model for the large-field inflation}
\label{sec:linflation}
In this section, we discuss how to realize the 
large-field inflation that would explain the WMAP, 
Planck~\cite{Ade:2013zuv} and BICEP2 data~\cite{Ade:2014xna}, 
although there is a possible tension between these collaborations. 
 
Unlike the previous section, we consider two light pair of modulus 
and stabilizer fields, e.g. $T^{I'=1,2}$ and $H_{i=1,2}$ 
which are decoupled from the other heavy pairs $T^{I'\neq 1,2}$ and 
$H_{i\neq 1,2}$. This scenario can be realized 
when $c_{I'=1,2}^{(i \ne 1,2)}=c_{I' \ne 1,2}^{(i=1,2)}=0$ 
in Eq.~(\ref{eq:4dw}), $\left| c_{I' \ne 1,2}^{(i \ne 1,2)} \right|
<\left| c_{I'=1,2}^{(i=1,2)} \right|$ 
in Eq.~(\ref{eq:thmass}), $\left|J_{0}^{i= 1,2}\right| 
< \left| J_{0}^{i\neq 1,2}\right|$ and 
$\left|J_{L}^{i= 1,2}\right| 
< \left| J_{L}^{i\neq 1,2}\right|$ in Eq.~(\ref{eq:thmass}). 
Below the heavier mass scale $m_{I' \ne 1,2\,i\ne 1,2}$, 
the effective K\"ahler potential and superpotential for 
the light fields $T^{I'=1,2}$ and $H_{i=1,2}$ are given 
by 
\begin{eqnarray}
K_{\rm eff}(T^1,H_1,T^2,H_2) 
&=& -\ln {\cal N}({\rm Re}\,T)\,\Big|_0
+Z_{1,\bar{1}}
({\rm Re}\,T)\,\Big|_0 \,|H_1|^2
+Z_{2,\bar{2}}
({\rm Re}\,T)\,\Big|_0 \,|H_2|^2, 
\nonumber \\
W_{\rm eff}(T^1,H_1,T^2,H_2) 
&=& \left( J_0^{(1)} 
+e^{-c_{I'}^{(1)}T^{I'}}J_L^{(1)} \right) H_1 
+\left( J_0^{(2)} 
+e^{-c_{I'}^{(2)}T^{I'}}J_L^{(2)} \right) H_2, 
\label{eq:weff2}
\end{eqnarray}
where 
$f(T^{I'},H_i)\,\Big|_0 \equiv f(T^{I'},H_i)
\Big|_{\scriptsize \begin{array}{ll}
T^{I' \ne 1,2}=\langle T^{I' \ne 1,2} \rangle \\ 
H_{i \ne 1,2}=\langle H_{i \ne 1,2} \rangle
\end{array}}$ 
for an arbitrary function 
$f(T^{I'},H_i)$, and then 
\begin{eqnarray}
Z_{1,\bar{1}}({\rm Re}\,T)\,\Big|_0 = 
\frac{1-e^{-2c_{I'}^{(1)}{\rm Re}\,T^{I'}}}{
c_{I'}^{(1)} {\rm Re}\,T^{I'}}, \,\,\,\,
Z_{2,\bar{2}}({\rm Re}\,T)\,\Big|_0 = 
\frac{1-e^{-2c_{I'}^{(2)}{\rm Re}\,T^{I'}}}{
c_{I'}^{(2)} {\rm Re}\,T^{I'}}. 
\label{eq:kml}
\end{eqnarray}
The effective potential for the light fields, 
\begin{eqnarray}
V_{\rm eff}(T^1,H_1,T^2,H_2) &=& e^{K_{\rm eff}} 
\left( (K_{\rm eff})^{m,\bar{n}}\,
D_mW_{\rm eff}\,D_{\bar{n}} \bar{W}_{\rm eff}
-3|W_{\rm eff}|^2 \right), 
\label{eq:4dspeff2}
\end{eqnarray}
is obtained by using the effective K\"ahler potential 
and superpotential~(\ref{eq:weff2}), 
where $m,n=\{I',i\}$ with $I'=1,2$ and $i=1,2$. 

The most general form of part of norm function 
carrying the two light moduli $T^1$ and $T^2$ is written as 
\begin{eqnarray}
{\cal N} ({\rm Re} T)\Big|_0 &=& C_{1,1,1}({\rm Re} T^1)^3 + 
C_{1,1,2}({\rm Re} T^1)^2({\rm Re} T^2) 
+C_{1,2,2} ({\rm Re} T^1)({\rm Re} T^2)^2 
+C_{2,2,2} ({\rm Re} T^2)^3 \nonumber\\
&+& \cdots,
\label{eq:gnorm}
\end{eqnarray}
where the ellipsis stands for terms those do not contain 
the two light moduli.\footnote{
We assume the couplings between the decoupled fields 
$T^{I'}$ with $I^{'}=3,4,\cdots$ and the lighter fields 
$T^1$ and $T^2$ are absent in the norm function for simplicity.} 
To brighten the outlook for analyzing the 
above scalar potential (\ref{eq:4dspeff2}),
 we redefine the modulus field as 
\begin{equation}
\;\hat{T^1} \equiv \frac{c_1^{(1)} T^1+c_2^{(1)} T^2}{c},\;\;\;
\;\hat{T^2} \equiv \frac{c_1^{(2)} T^1+c_2^{(2)} T^2}{d}.
\end{equation}
where $c$ and $d$ are the $U(1)$ charge of the stabilizer field 
$H_1$ and $H_2$ for a linear combination of the $Z_2$-odd vector 
fields $A_M^{I'}$ in ${\bf V}^{I'}$ with $I'=1,2$, respectively. 
In this field base $\hat{T}^1$ and $\hat{T}^2$, the 
mixing terms between $\hat{T}^1$ and $\hat{T}^2$ in the 
superpotential are canceled and thus each of 
$\hat{T}^1$ and $\hat{T}^2$ has the independent 
superpotential to each other, 
\begin{equation}
W_{\rm eff}(\hat{T}^1,H_1,\hat{T}^2,H_2) 
\,=\, \left( J_0^{(1)} 
+e^{-c\,\hat{T}^1}J_L^{(1)} \right) H_1 
+\left( J_0^{(2)} 
+e^{-d\,\hat{T}^2}J_L^{(2)} \right) H_2. 
\label{eq:resup}
\end{equation}
The vacuum expectation values of moduli $\hat{T}^1, 
\hat{T}^2$ and stabilizer fields $H_1, H_2$ are determined by 
minimizing the scalar 
potential~(\ref{eq:4dspeff2}) in a similar way to those of the 
small-field inflation (\ref{eq:movev}) as 
\begin{eqnarray}
\langle \hat{T}^{1} \rangle 
&=& \frac{1}{c} \ln \frac{J_L^{(1)}}{J_0^{(1)}}, \qquad 
\langle \hat{T}^{2} \rangle \ = \ 
\frac{1}{d} \ln \frac{J_L^{(2)}}{J_0^{(2)}}, \qquad 
\langle H_1 \rangle \ = \ \langle H_2 \rangle \ = \ 0, 
\label{eq:vacuum}
\end{eqnarray}
which satisfy 
$\langle D_{\hat{I}'}W \rangle 
= \langle D_iW \rangle 
= \langle W \rangle 
= 0$ 
and then 
$\langle V_{\hat{I}'} \rangle 
= \langle V_i \rangle 
= \langle V \rangle 
= 0$
for 
$V_m=\partial_m V$. 

When we construct the large-field inflation model in the 
next subsection, we restrict ourselves to the case that 
the coefficients $C_{I',J',K'}$ and the charges $c_{I'}^{(i)}$ 
for $I',J',K'=1,2$ and $i=1,2$ are chosen in such a way 
that the norm function is written as 
\begin{equation}
{\cal N} ({\rm Re} T)\Big|_0= a({\rm Re} \hat{T}^1)
({\rm Re} \hat{T}^2 -b\,{\rm Re} \hat{T}^1)^2 +\cdots,
\label{eq:norm}
\end{equation}
in the hatted field base, where $a$ and $b$ are positive 
real numbers determined by the fixed values of $C_{I',J',K'}$ 
and $c_{I'}^{(i)}$ as, e.g., 
$a=c_1^{(1)}(d)^2/c\,(c_2^{(2)})^2,b=c\,c_1^{(2)}/d\,c_1^{(1)}$ 
for 
$c_2^{(1)}=C_{1,1,1}=C_{1,1,2}=C_{2,2,2}=0$ and $C_{1,2,2}=1$. 
The ellipsis in Eq.~(\ref{eq:norm}) has same meaning as that 
in Eq.~(\ref{eq:gnorm}) and is irrelevant in the following 
arguments. 
The advantage of the norm function (\ref{eq:norm}) will be 
explained in Sec.~\ref{subsec:linflation}, and here we 
notice that it leads to the moduli 
mixing in the K\"ahler metric, $K_{\hat{I}^{'},\bar{\hat{J}}^{'}} 
\neq 0$ for $\hat{I}^{'} \neq \hat{J}^{'}$. 
We have to check the positivity of the Hessian matrix with 
the scalar potential (\ref{eq:4dspeff2}). The mass 
matrix given by the potential (\ref{eq:4dspeff2}) 
is written in a block-diagonal form with two nonvanishing 
blocks according to the absence 
mixing terms between the moduli $\hat{T}^{I'=1,2}$ 
and the stabilizer fields $H_{i=1,2}$. 
Since the following mixing terms are all vanishing at 
the vacuum, 
\begin{eqnarray}
&\langle V_{\hat{T}^1\bar{\hat{T}}^2} \rangle =
\langle V_{\hat{T}^1\bar{H}_1} \rangle =
\langle V_{\hat{T}^1\bar{H}_2} \rangle =
\langle V_{\hat{T}^2\bar{H}_1} \rangle =
\langle V_{\hat{T}^2\bar{H}_2}\rangle =0,\nonumber\\
&\langle (K_{\rm eff})^{\hat{T}^1\bar{H}_1} \rangle =
\langle (K_{\rm eff})^{\hat{T}^1\bar{H}_2} \rangle =
\langle (K_{\rm eff})^{\hat{T}^2\bar{H}_1} \rangle =
\langle (K_{\rm eff})^{\hat{T}^2\bar{H}_2} \rangle =0, 
\end{eqnarray}
for $V_{mn}=\partial_n\partial_m V$, 
there are no mixing between the moduli $\hat{T}^{I'}$ and the stabilizer 
fields $H_i$ at the vacuum, that is, the mass (sub-)matrices of 
the moduli $\hat{T}^{I'}$ and the stabilizer fields $H_i$ can be 
analyzed independently. 

First, we consider the mass-squared matrix $m_{t}^2$ of the 
real parts of the moduli in the base of canonically normalized 
field $t^{I'}$, 
\begin{equation}
t^{I'} \,=\,  \sum_{J'=1}^2 \sqrt{2(K_{\hat{T}})_{I'}} U_{I',J'}
{\rm Re}\,\hat{T}^{J'},
\label{eq:canbas}
\end{equation}
for $I'=1,2$, which is estimated as 
\begin{equation}
m_t^2 =
\begin{pmatrix}
\sqrt{\frac{1}{(K_{\hat{T}})_1}} & 0 \\
0 & \sqrt{\frac{1}{(K_{\hat{T}})_2}} 
\end{pmatrix}
U
\begin{pmatrix}
V_{\hat{T}^1\bar{\hat{T}}^1} & 0 \\
0 & V_{\hat{T}^2\bar{\hat{T}}^2}
\end{pmatrix}
U^{-1}
\begin{pmatrix}
\sqrt{\frac{1}{(K_{\hat{T}})_1}} & 0 \\
0 & \sqrt{\frac{1}{(K_{\hat{T}})_2}} 
\end{pmatrix}
,
\label{eq:masst}
\end{equation}
where $V_{\hat{T}^{I'}\bar{\hat{T}}^{J'}} =
\langle 
e^{K_{\rm eff}} (K_{\rm eff})^{H_i\bar{H}_j}
W_{\hat{T}^{I'}H_i}\overline{W_{\hat{T}^{J'}H_j}} 
\rangle$, and
$(K_{\hat{T}})_1$, $(K_{\hat{T}})_2$ and $U$ are the 
eigenvalues and diagonalizing matrix of the moduli 
K\"ahler metric, respectively. 
(Explicit form of them are shown in 
Appendix \ref{app:can}.) 
By contrast, the mass-squared matrix 
of the imaginary part, 
${\rm Im}\,\hat{T}^1$ and ${\rm Im}\,\hat{T}^2$, is 
already diagonalized because K\"ahler potential 
does not contain the imaginary parts of moduli, those 
are prohibited by the $U(1)$ gauge symmetries. The 
supersymmetric masses of canonically normalized moduli 
$\phi^{I'}$, 
\begin{equation}
\phi^{I'} \,=\, \sqrt{2(K_{\rm eff})_{\hat{T}^{I'}\bar{\hat{T}}^{I'}}}\,{\rm Im}\,\hat{T}^{I'},
\label{eq:canbasphi}
\end{equation}
for $I'=1,2$, are also same as those shown 
in Eq. (\ref{eq:thmass}) where the K\"ahler potential and 
superpotential are replaced by Eq. (\ref{eq:weff2}).

Second, the mass-squared matrix 
of the stabilizer field is also evaluated 
in the base of canonically normalized field $h_i$,
\begin{equation}
h_i \,=\, \sum_{j=1}^2 \sqrt{2(K_{H})_i} 
\delta_{i,j} H_j,
\label{eq:canbash}
\end{equation}
for $i=1,2$, that is found as 
\begin{equation}
m_h^2 =
\begin{pmatrix}
\sqrt{\frac{1}{(K_H)_1}} & 0 \\
0 & \sqrt{\frac{1}{(K_H)_2}} 
\end{pmatrix}
\begin{pmatrix}
V_{H_1\bar{H}_1} & V_{H_1\bar{H}_2} \\
V_{H_2\bar{H}_1} & V_{H_2\bar{H}_2}
\end{pmatrix}
\begin{pmatrix}
\sqrt{\frac{1}{(K_H)_1}} & 0 \\
0 & \sqrt{\frac{1}{(K_H)_2}} 
\end{pmatrix}
,
\label{eq:massh}
\end{equation}
where $V_{H_i\bar{H}_j} =\langle 
e^{K_{\rm eff}} 
(K_{\rm eff})^{\hat{T}^{I'}\bar{\hat{T}}^{J'}}
W_{\hat{T}^{I'}H_i}\overline{W_{\hat{T}^{J'}H_j}} 
\rangle$, and 
$(K_H)_1 =\langle (K_{\rm eff})_{H_1\bar{H}_1} \rangle$, 
$(K_H)_2 =\langle (K_{\rm eff})_{H_2\bar{H}_2} \rangle$ 
are the eigenvalues of the K\"ahler metric 
of the stabilizer fields, respectively. 

From the mass-squared 
matrices (\ref{eq:masst}) and 
(\ref{eq:massh}), the supersymmetric masses 
of the canonically normalized stabilizer fields 
$h_i$ and the moduli $t^{I'}$ in the limit of 
$\langle W_{\hat{T}^1H_1}\rangle \ll
\langle W_{\hat{T}^2H_2}\rangle$ are estimated as, 
\begin{align}
&m_{t^1}^2 \simeq
\frac{e^{\langle K_{\rm eff} \rangle} 
\langle (K_{\rm eff})^{H_1\bar{H}_1}\rangle 
\langle W_{\hat{T}^1H_1}\rangle^2}
{\langle (K_{\rm eff})_{\hat{T}_1\bar{\hat{T}}_1}\rangle}, 
\qquad
m_{t^2}^2 \simeq
e^{\langle K_{\rm eff} \rangle} 
\langle (K_{\rm eff})^{H_2H_2}\rangle 
\langle (K_{\rm eff})^{\hat{T}^2\bar{\hat{T}}^2}\rangle
\langle W_{\hat{T}^2H_2}\rangle^2, 
\nonumber\\
&m_{{\rm Re}\,h_1}^2=m_{{\rm Im}\,h_1}^2\simeq
\frac{e^{\langle K_{\rm eff} \rangle} 
\langle (K_{\rm eff})^{\hat{T}^1\bar{\hat{T}}^1}\rangle 
\langle W_{\hat{T}^1H_1}\rangle^2}
{\langle (K_{\rm eff})_{H_1\bar{H}_1}\rangle}, \nonumber\\
&m_{{\rm Re}\,h_2}^2=m_{{\rm Im}\,h_2}^2\simeq
\frac{e^{\langle K_{\rm eff} \rangle} 
\langle (K_{\rm eff})^{\hat{T}^2\bar{\hat{T}}^2}\rangle 
\langle W_{\hat{T}^2H_2}\rangle^2}
{\langle (K_{\rm eff})_{H_2\bar{H}_2}\rangle}. 
\label{eq:massd}
\end{align}
These expressions show that the squared 
masses of moduli and stabilizers are all positive 
at the vacuum if there is a hierarchy 
$\langle W_{\hat{T}^1H_1}\rangle \ll
\langle W_{\hat{T}^2H_2}\rangle$ as mentioned 
in Sec. \ref{sec:moduli}, that confirms the stability 
of the vacuum~(\ref{eq:vacuum}). \footnote{If the hierarchy 
$\langle W_{\hat{T}^1H_1}\rangle \ll
\langle W_{\hat{T}^2H_2}\rangle$ does not 
exist, sizable K\"ahler mixings may spoil the 
stability of the vacuum.} 

Because the two pair $(\hat{T}^1,H_1)$ and $(\hat{T}^2,H_2)$ 
of modulus and stabilizer have totally independent vacuum 
expectation values to each other as shown in 
Eq.~(\ref{eq:vacuum}), we can further consider the situation 
that the first pair ($\hat{T}^1,H_1$) is lighter than the second 
pair ($\hat{T}^2,H_2$) by assuming 
$\left|J_0^1\right| < \left|J_0^2\right|$ and 
$\left|J_L^1\right| < \left|J_L^2\right|$. 
In this case, the second pair can also be integrated out, and 
the effective potential for the first pair is given by
\begin{eqnarray}
V_{\rm eff}(\hat{T}^1,H_1) &=& e^{K_{\rm eff}} 
\left( (K_{\rm eff})^{m,\bar{n}}\,
D_mW_{\rm eff}\,D_{\bar{n}} \bar{W}_{\rm eff}
-3|W_{\rm eff}|^2 \right), 
\label{eq:veffl}
\end{eqnarray}
where $m,n=\{\hat{I}',i\}$ with $\hat{I}'=1$ and $i=1$ 
and the effective K\"ahler potential $K_{\rm eff}$
and superpotential $W_{\rm eff}$ are obtained as
\begin{eqnarray}
K_{\rm eff}(\hat{T}^1,H_1) 
&=& -\ln {\cal N}({\rm Re}\,\hat{T})\,\Big|_0
+Z_{1,\bar{1}}
({\rm Re}\,\hat{T})\,\Big|_0 \,|H_1|^2, 
\nonumber \\
W_{\rm eff}(\hat{T}^1,H_1) 
&=& \left( J_0^{(1)} 
+e^{-c\,\hat{T}^{1}}J_L^{(1)} \right) H_1. 
\label{eq:weff3}
\end{eqnarray}
Here we adopt the notation 
$f(T^{I'},H_i)\,\Big|_0 \equiv f(T^{I'},H_i)
\Big|_{\scriptsize \begin{array}{lll}
T^{I' \ne 1,2}=\langle T^{I' \ne 1,2} \rangle \\
\hat{T}^{I' \ne 2}=\langle \hat{T}^{I' \ne 2} \rangle \\ 
H_{i \ne 1}=\langle H_{i \ne 1} \rangle \\
\end{array}}$ 
for an arbitrary function 
$f(T^{I'},H_i)$.

\subsection{The Inflation potential and dynamics}
\label{subsec:linflation}
\begin{figure}[t]
\centering \leavevmode
\includegraphics[width=0.5\linewidth]{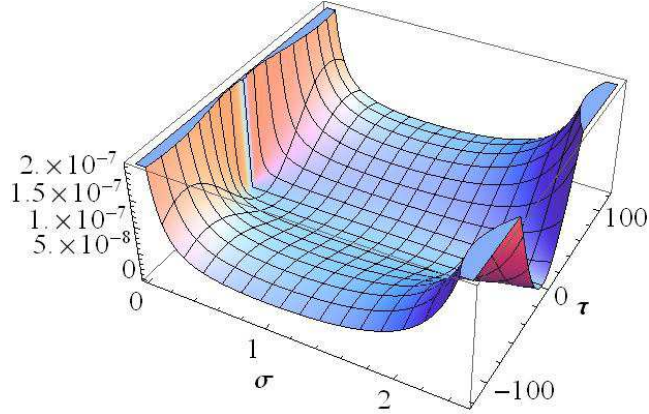}
\caption{The scalar potential~(\ref{eq:veff2}) on the 
$(\sigma, \tau)$-plane.}
\label{fig:linfpo}
\end{figure}

\begin{figure}[t]
\centering \leavevmode
\includegraphics[width=0.5\linewidth]{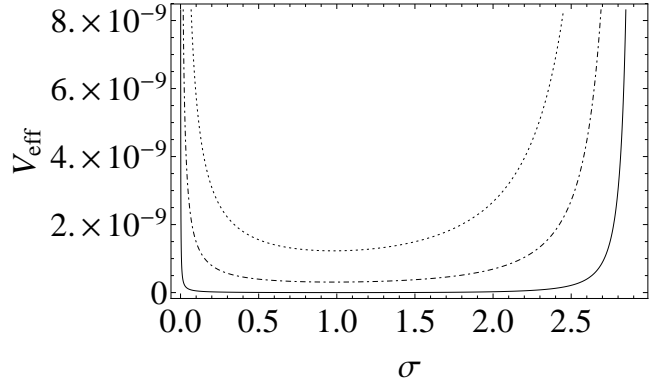}
\caption{The scalar potential~(\ref{eq:veff2}) on the hypersurfaces $\tau=10$ 
(dotdashed line), $\tau=5$ (dashed line) and $\tau=0$ 
(thick line).}
\label{fig:linfReT1}
\end{figure}

From the above scalar potential (\ref{eq:veffl}), 
we find the effective potential for the modulus 
$\hat{T}^1$ on the $H_1 =0$ 
hypersurface,
\begin{eqnarray}
V_{\rm eff}(\hat{T}^1,H_1=0) 
&=& e^{K_{\rm eff}}
(K_{\rm eff})^{i=1,\bar{i}=\bar{1}} 
\left| (W_{\rm eff})_{i=1} \right|^2 
\,\Big|_{H_1=0}
\nonumber \\ &=& 
\Lambda^4 (1-\lambda\,\text{cos}(c\,\tau )), 
\label{eq:veff2}
\end{eqnarray}
where
\begin{eqnarray}
\Lambda^4 &\equiv & 
\frac{c}{(\langle {\rm Re}\,\hat{T}^2\rangle -b\, \sigma)^2}
\frac{J_{01}^2 + J_{L1}^2 e^{-2c\,\sigma}}{1-e^{-2c\,\sigma}},
\nonumber \\
\lambda &\equiv & 2\cfrac{J_{01}J_{L1}e^{-c\,\sigma}}{J_{01}^2 
+ J_{L1}^2 e^{-2c\,\sigma}},
\label{eq:deflambda}
\end{eqnarray}
we adopted the norm function (\ref{eq:norm}) 
with $a=1$ and $\hat{T}^1 = \sigma + i\tau$. 
Fig. \ref{fig:linfpo} shows the scalar potential on the 
$(\sigma, \tau)$-plane, where the parameters 
are chosen as 
\begin{equation}
c=1/30,\qquad J_L^{(1)}\,=\,-4.4\times 10^{-3},
\qquad J_0^{(1)}\,=\,4.25\times 10^{-3},\qquad b=15,
\label{parameter}
\end{equation}
in the Planck unit $M_{{\rm Pl}} =1$. 
The imaginary direction $\tau$ has a periodic 
property as can be seen in Eq.~(\ref{eq:veff2}) and 
the real direction $\sigma$ will be stabilized at the minimum 
shown in Fig. \ref{fig:linfReT1} in which 
the behaviors of potential on the hypersurfaces $\tau=10$ 
(dot dashed line), $\tau=5$ (dashed line) and 
$\tau=0$ (thick line) are drawn. 
As we can see from Fig. \ref{fig:linfReT1}, the negative 
region of $\sigma<0$ is not allowed, because the 
$\Lambda$ in Eq. (\ref{eq:veff2}) diverges in the 
limit of $\sigma \rightarrow 0$, while the overshooting 
to a large-field region, $\sigma >$ 
$\langle {\rm Re}\,\hat{T}^2 \rangle /b$, 
is also prohibited by the structure of the 
norm function (\ref{eq:norm}). 
Since Re\,$\hat{T}^2$ is already stabilized by its own 
minimum (\ref{eq:vacuum}), we find 
\begin{equation}
\lim_{{\rm Re}\,\hat{T}^1 \to \langle {\rm Re}\,\hat{T}^2
\rangle /b} 
\left| V_{\rm eff}(\hat{T}^1,H_1=0) \right| 
= \infty.  
\end{equation}

From these properties of the potential~(\ref{eq:veff2}), 
we expect that the so-called natural inflation~\cite{Freese:1990rb} 
would occur by identifying $\tau={\rm Im}\,\hat{T}^1$ 
as the inflaton field. 
During the inflation, the real part $\sigma$ will take 
a different field value from the one at the true minimum~(\ref{eq:vacuum}) 
and after the inflation, it rolls down to the minimum and oscillates around 
the vacuum. 
To confirm the above statements, we solve the equations of 
motion for two fields $\sigma$ and $\tau$ under the assumption 
that the oscillations of the stabilizer fields, Re\,$H_1$ and Im\,$H_1$, 
around their vacuum expectation values $\langle {\rm Re}\,H_1 
\rangle =\langle {\rm Im}\,H_1 \rangle =0$ are negligible 
during and after the inflation (which will be confirmed 
in Sec. \ref{subsec:modpro}). The equations of motion for these 
fields are written as 
\begin{align}
\sigma^{''} &= -\left(1-{\cal L}_{\text{kin}} \right)
\left( 3\sigma^{'} +6\frac{\sigma^2 (\langle \hat{T}^2\rangle -b\sigma)^2}{(\langle \hat{T}^2\rangle -b\sigma)^2 +2b^2\sigma^2}\frac{\partial_{\sigma}V}{V}\right) +\frac{ (\sigma^{'})^2 -(\tau^{'})^2}{\sigma (\langle \hat{T}^2\rangle -b\sigma)} \left( \frac{(\langle \hat{T}^2\rangle -b\sigma)^3 -2b^3\sigma^3}{(\langle \hat{T}^2\rangle -b\sigma)^2 +2b^2\sigma^2}\right), \nonumber\\
\tau^{''} &= -\left(1-{\cal L}_{\text{kin}} \right)\left( 3\tau^{'} +6\frac{\sigma^2 (\langle \hat{T}^2\rangle -b\sigma)^2}{(\langle \hat{T}^2\rangle -b\sigma)^2 +2b^2\sigma^2}\frac{\partial_{\tau}V}{V}\right) +\frac{ 2\sigma^{'}\tau^{'}}{\sigma (\langle \hat{T}^2\rangle -b\sigma)} \left( \frac{(\langle \hat{T}^2\rangle -b\sigma)^3 -2b^3\sigma^3}{(\langle \hat{T}^2\rangle -b\sigma)^2 +2b^2\sigma^2}\right), \nonumber\\
{\cal L}_{\text{kin}} &\equiv\frac{(\langle \hat{T}^2\rangle -b\sigma)^2 +2b^2\sigma^2}{2\sigma^2 (\langle \hat{T}^2\rangle -b\sigma)^2} \left((\sigma^{'})^2 +(\tau^{'})^2\right),
\label{eq:eqm}
\end{align}
where the prime denotes the derivative $d/dN$ with respect 
to the number $N$ of e-foldings as before, and we described 
the Christoffel symbol for the target space in terms of the metric  $g_{\sigma\sigma}=g_{\tau\tau}=\frac{(\langle \hat{T}^2\rangle -b\sigma)^2 +2b^2\sigma^2}{2\sigma^2 (\langle \hat{T}^2\rangle -b\sigma)^2}$. 

In the following analysis, the numerical values of 
parameters in the Planck unit $M_{{\rm Pl}} = 1$ are chosen 
as 
\begin{equation}
d\,=\,1/20, \qquad J_L^{(2)}/J_0^{(2)}\,=\,-9,\qquad J_0^{(0)}\,=\,10^{-1},
\end{equation} 
for the heavier fields $\hat{T}^2$ and $H_2$ as well as 
those (\ref{parameter}) for the light fields $\hat{T}^1$ 
and $H_1$. With these parameters, the vacuum 
expectation values of fields are given by
\begin{equation}
\langle \hat{T}^1\rangle \simeq 1.04, \qquad 
\langle \hat{T}^2\rangle \,=\, 43.94, \qquad 
\langle H_1\rangle \,=\, \langle H_2\rangle \,=\,0.
\end{equation}
For the canonically normalized fields (\ref{eq:canbas}), (\ref{eq:canbasphi}) 
and (\ref{eq:canbash}), their vacuum expectation values are 
\begin{equation}
\langle t^1\rangle \simeq 2.29, \qquad 
\langle t^2\rangle \,\simeq\, 1.22, \qquad 
\langle \phi^1\rangle \,=\, \langle \phi^2\rangle 
\,=\,
\langle h_1\rangle \,=\, \langle h_2\rangle \,=\,0.
\end{equation}
The supersymmetric masses (\ref{eq:thmass}) 
and (\ref{eq:massd}) 
of these fields $t^{I'}$,$\phi^{I'}$ and $h_{i}$ 
are estimated as 
\begin{align}
&(m_{t^1})^2 \simeq (m_{\phi^1})^2 \simeq 
(m_{{\rm Re}\,h_1})^2 =(m_{{\rm Im}\,h_1})^2 
\simeq 
\left( 1.96 \times 10^{13}\,{\rm GeV}\right)^2, 
\nonumber\\ 
&(m_{t^2})^2 \simeq 
(m_{{\rm Re}\,h_2})^2 =(m_{{\rm Im}\,h_2})^2 
\simeq 
\left( 4.45 \times 10^{16}\,{\rm GeV}\right)^2, 
\nonumber\\
&(m_{\phi^2})^2 
\simeq 
\left( 3.5 \times 10^{16}\,{\rm GeV}\right)^2. 
\label{eq:massvac}
\end{align}
Note that the supersymmetric masses of $\phi^{I'}$ are 
in general different from those of $t^{I'}$ due to the different 
canonical normalization of Re$\,\hat{T}^{I'}$ 
and Im$\,\hat{T}^{I'}$ from each other. The Hubble scale is given by 
\begin{equation}
H_{\rm inf}\,=\,\left( V_{\rm inf}/3M_{\rm Pl}^2\right)^{1/2} 
\simeq 8.6 \times 10^{13} {\rm GeV},
\label{eq:hubl}
\end{equation}
where $V_{\rm inf} \sim \Lambda^4$ is estimated by Eq. (\ref{eq:veff2}). 
We check that these masses (\ref{eq:massvac}) 
and Hubble scale (\ref{eq:hubl}) are below the 
compactification scale 
$M_C \simeq 
\pi M_{\rm Pl}/\langle 
{\cal N}({\rm Re}\,\hat{T}) \rangle^{1/2} 
\simeq 2.6 \times 10^{17} {\rm GeV}$ 
to ensure the validity of 4D effective-theory 
description. 
The pair ($\hat{T}^2,H_2$) 
is stabilized at $\hat{T}^1$- and $H_1$-independent 
minimum and their masses are larger than the inflaton scale, 
that is, $H_{\rm inf}^2\ll 
(m_{t^2}^2),\,(m_{\phi^2}^2),\,(m_{{\rm Re}\,h^2}^2),\,(m_{{\rm Im}\,h^2}^2)$. 
Then the heavier pair ($\hat{T}^2,H_2$) is decoupled 
from the inflation dynamics.

Next we define the slow-roll 
parameters for the multi-field case~\cite{Burgess:2004kv} 
to estimate the observable quantities constrained by the 
cosmological observations,
 \begin{align}
\epsilon &
=\,\frac{g^{\sigma\sigma}}{2}\left(
\frac{\partial_{\sigma} V}{V}\right)^2 
+\frac{g^{\tau\tau}}{2}\left(
\frac{\partial_{\tau} V}{V}\right)^2, \nonumber\\
\eta &=\,\text{minimum eigenvalue of} \;\left \{
\frac{1}{V}
\begin{pmatrix}
\nabla^{i} \nabla_{j} V  & \nabla^{i} \nabla_{\bar{j}} V \\
\nabla^{\bar{i}} \nabla_{j} V & \nabla^{\bar{i}} \nabla_{\bar{j}} V 
\end{pmatrix}
\right \}
,\nonumber\\
&=\frac{g^{\sigma\sigma}}{2}
\left(\frac{\partial_{\sigma} \partial_{\sigma} V}{V}
+\frac{\partial_\tau \partial_\tau V}{V} 
-\sqrt{\left(\frac{\partial_{\sigma} \partial_{\sigma} V}{V} 
-\frac{\partial_{\tau} \partial_{\tau} V}{V} 
-2\Gamma^{\sigma}_{\sigma\sigma} 
\frac{\partial_{\sigma} V}{V} \right)^2 
+4\left( \frac{\partial_{\sigma} \partial_{\tau} V}{V} -
\Gamma^{\sigma}_{\sigma\sigma} \frac{\partial_{\tau} V}{V} \right)^2} \right),
\end{align}
where $i,j\,=\,\sigma,\tau$. 
The observables such as the power spectrum $P_\xi (k)$ 
of scalar curvature perturbation, its spectral index $n_s$ 
and the tensor-to-scalar ratio $r$ 
are written in terms of these slow-roll parameters, 
\begin{align}
P_\xi (k)&=\frac{1}{24\pi^2} \frac{V}{\epsilon\, M_{Pl}^4}, 
\nonumber\\
n_s &= 1+ \frac{d\,{\rm ln} P_{\xi}(k)}{d\,{\rm ln}\,k} 
\simeq 1-6\epsilon +2\eta, \nonumber\\
r &=16\,\epsilon.
\label{eq:obs}
\end{align}

\begin{figure}[t]
\begin{minipage}{0.5\hsize}
\begin{center}
\includegraphics[width=0.8\linewidth]{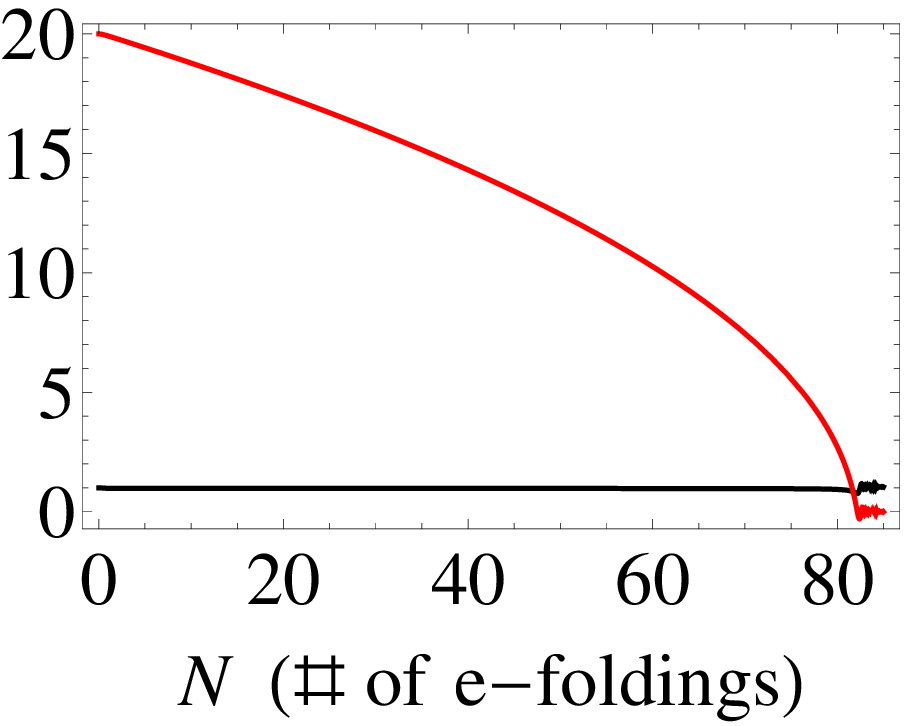}
\end{center}
\end{minipage}
\begin{minipage}{0.5\hsize}
\begin{center}
\includegraphics[width=0.8\linewidth]{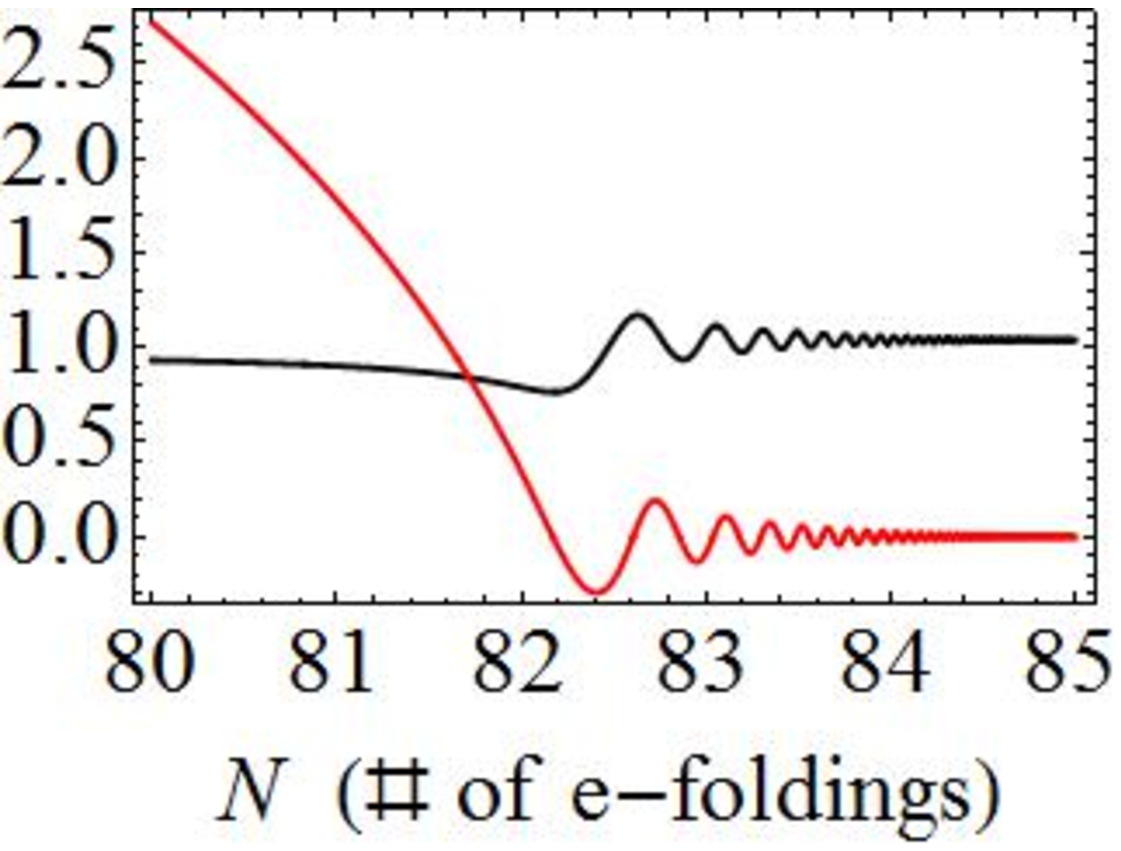}
\end{center}
\end{minipage}
\caption{The behavior of the $\sigma={\rm Re}\,\hat{T}^1$ 
(black curves) and $\tau={\rm Im}\,\hat{T}^1$ (red curves) 
as a function of the e-folding number $N$.}
\label{fig:linfosc}
\end{figure}

We numerically solve Eq. (\ref{eq:eqm}) with the initial 
conditions $(\sigma, \tau)=(1,20)$ and 
$(\sigma^{'}, \tau^{'})=(0,0)$ at $N=0$ and then 
Fig. \ref{fig:linfosc} 
shows the evolution of $\sigma$ 
and $\tau$ as a function of $N$. The time at 
the end of inflation corresponds to an about 
$N_{\rm end}\simeq 81.2$ e-folds, when the slow-roll 
condition is violated (max\,$\{\epsilon, \eta\} =1$). 
Fig. \ref{fig:linfosc} confirms a desired situation 
that the real part $\sigma$ of the light modulus 
is fixed to a certain field value different from its 
vacuum expectation value during the inflation 
and oscillates around the vacuum after the inflation. 
Such a dynamics is explained as follows. 
The mass square of $\sigma$ consists of those from 
the Hubble-induced and the supersymmetric 
contributions, 
\begin{align}
\partial_\sigma\partial_\sigma V &\simeq \,
3f(\sigma) H_{\rm inf}^2 + m_{\rm SUSY}^2,
\label{eq:Ret1mass}
\end{align}
where $H_{\rm inf}$ 
is the inflation scale 
defined by Eq. (\ref{eq:hubl}), 
$m_{\rm SUSY} \sim m_{t^1} 
\sim {\cal O}(10^{13}\,{\rm GeV})$ 
is the supersymmetric mass term originating 
from the superpotential (\ref{eq:weff2}) and 
$f(\sigma)$ is a function of $\sigma$ 
whose numerical value is of ${\cal O}(1)$ during and after the 
inflation. By virtue of the Hubble-induced 
contribution in Eq.~(\ref{eq:Ret1mass}), the 
real part $\sigma$ is ``stabilized" (at a different 
point from the minimum of potential) with 
its field value estimated below during the 
inflation caused by the slowly rolling imaginary 
part $\tau$ playing a role of inflaton field. 

\begin{figure}[t]
\begin{minipage}{0.5\hsize}
\begin{center}
\includegraphics[width=0.8\linewidth]{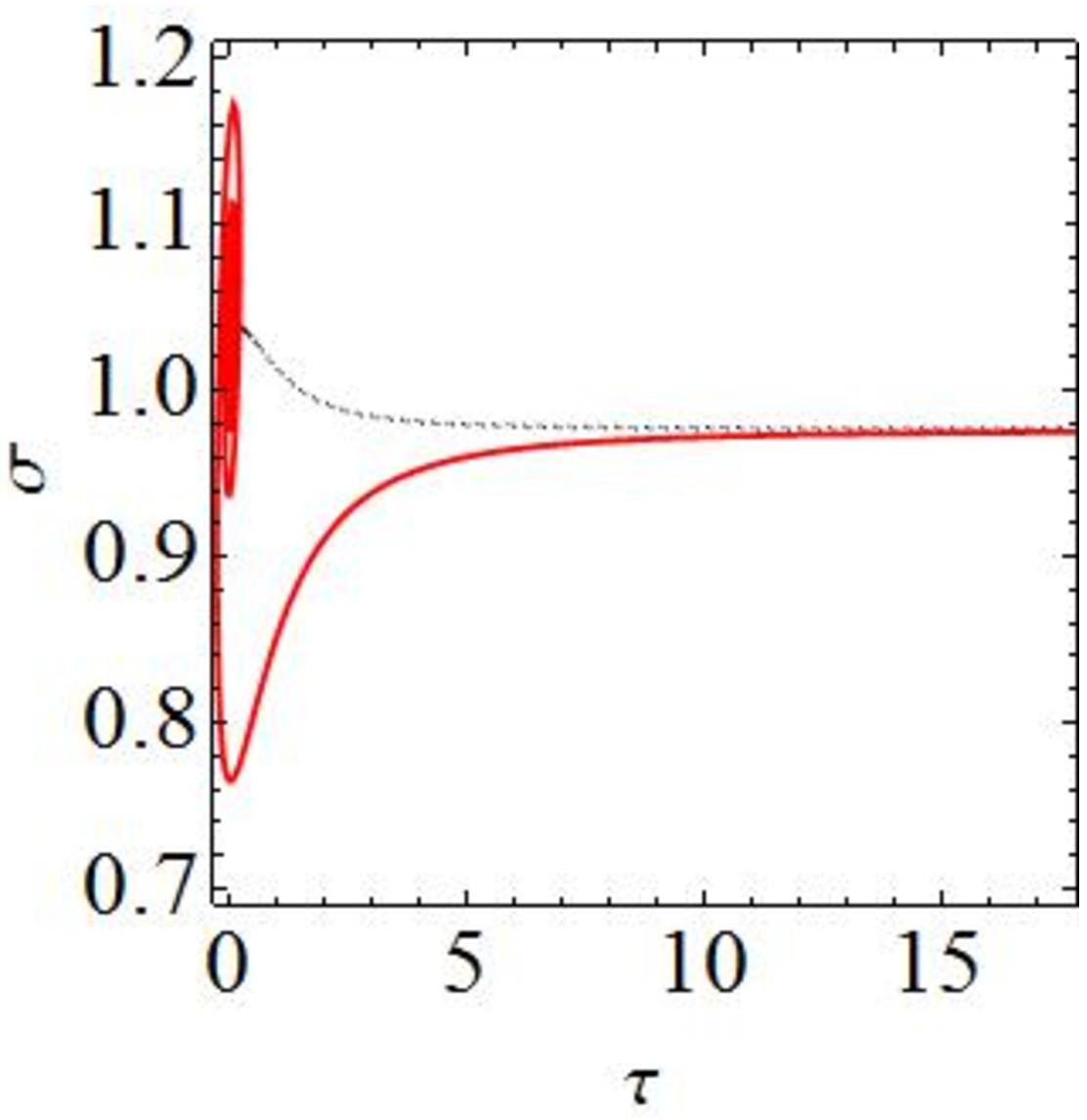}
\end{center}
\end{minipage}
\begin{minipage}{0.5\hsize}
\begin{center}
\includegraphics[width=0.8\linewidth]{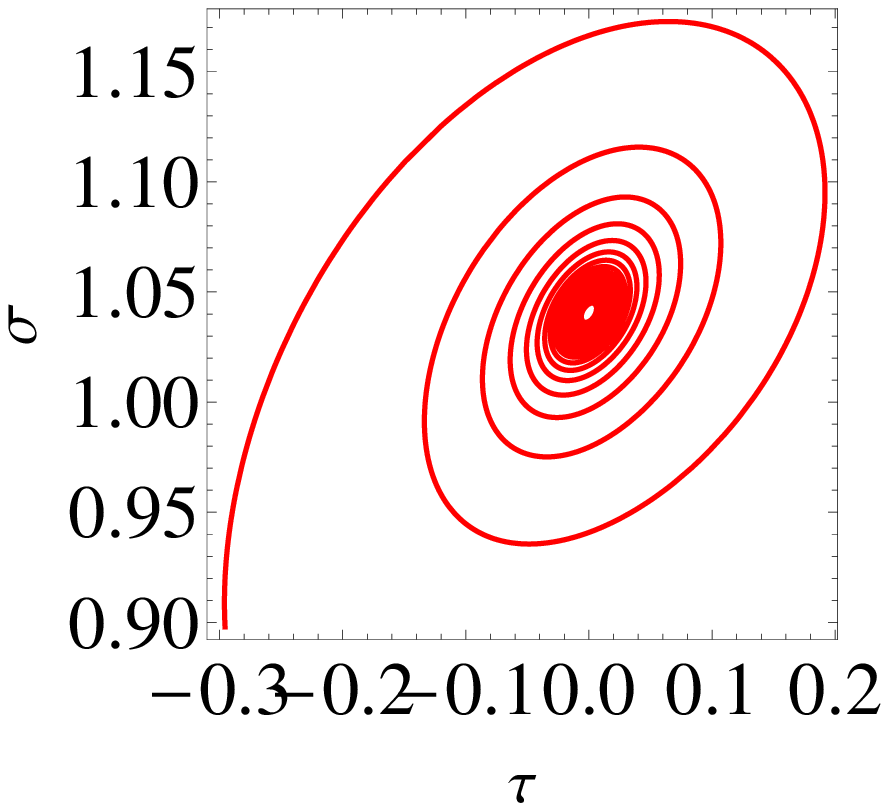}
\end{center}
\end{minipage}
\caption{The inflaton trajectory 
on the ($\tau-\sigma$)-plane. 
(The black dotted curve is evaluated under 
the slow-roll approximation (\ref{eq:trajsigma}), 
while the red solid curve is drawn by solving the 
full equations of motion (\ref{eq:eqm}) numerically.)}
\label{fig:linftra}
\end{figure}

The ``stabilized" value of $\sigma$ during the inflation 
can be estimated analytically from the approximated 
equation of motion for $\sigma$ under 
the slow-roll regime, $\sigma^{'}\ll 1$ and 
$\tau^{'}\ll 1$,
\begin{align}
\sigma^{'} &=-g^{\sigma\sigma}\frac{V_{\sigma}}{V} 
\nonumber \\
&=-g^{\sigma\sigma}\left( \frac{2}{
\langle \hat{T}^2 \rangle/b -\,\sigma} -
\frac{2c\,e^{-2c\,\sigma}}{1-e^{-2c\,\sigma}} 
-c \right)
+\frac{V_{\rm vac}(\sigma)}{V},
\label{eq:trajsigma}
\end{align}
where $V_{\rm vac}(\sigma) =e^{K_{\rm eff}} 
(K_{\rm eff})^{H_1\bar{H}_1}c
\left(\left|J_0^{(1)}\right|^2 
-\left|J_L^{(1)}\right|^2 e^{-2c\,\sigma} \right)
/{\cal N}$ 
and we dropped the mixing term proportional 
to $\sigma^{'} \tau^{'}$. 
The field value $\sigma=\sigma_{\rm inf}$ during the inflation 
is given by equaling the first parenthesis of Eq. 
(\ref{eq:trajsigma}) to $0$,
\begin{align}
&\frac{2}{
\langle \hat{T}^2 \rangle/b -\,\sigma_{\rm inf}} -
\frac{2c\,e^{-2c\,\sigma_{\rm inf}}}{1-e^{-2c\,\sigma_{\rm inf}}} 
-c \,=\,0, \nonumber\\
\Leftrightarrow 
&\frac{\langle \hat{T}^2 \rangle}{b} \,=\, 
\frac{-2+c\,\sigma_{\rm inf} 
+e^{2c\,\sigma_{\rm inf}}(2+c\,\sigma_{\rm inf})}
{c\,(1+e^{2c\,\sigma_{\rm inf}})}.
\end{align}
One of the advantages of the current setup in 
our model building is that we can choose 
the value of $\sigma_{\rm inf}$ close to the vacuum 
expectation value $\langle \sigma \rangle$ at the minimum 
of potential given by Eq. (\ref{eq:vacuum}), if we employ 
the parameters of the heavier modulus $\hat{T}^2$ 
in such a way that the following relation holds, 
\begin{equation}
\frac{\langle \hat{T}^2 \rangle}{b} \,\simeq\, 
\frac{-2+c\,\langle \sigma \rangle 
+e^{2c\,\langle \sigma \rangle}(2+c\,\langle \sigma \rangle)}
{c\,(1+e^{2c\,\langle \sigma \rangle})},
\label{eq:parab}
\end{equation}
which is already adopted in the above numerical 
analysis. 
Therefore, the inflaton dynamics caused by the light 
modulus field $\hat{T}^1=\sigma+i\tau$ can be 
dominated by its imaginary part $\tau$, and then it 
is classified as a single-field inflation which can avoid 
sizable magnitudes of the isocurvature 
perturbations possibly caused by the dynamics of the 
other fields (most likely $\sigma$) than the inflaton. 
As the inflaton $\tau$ rolls down toward the minimum 
(\ref{eq:vacuum}), the real part $\sigma$ also tends 
to go there,
because the value of $V$ approaches $V_{\rm vac}$ 
shown in Eq. (\ref{eq:trajsigma}). 
The discussion here is 
confirmed in Fig. \ref{fig:linftra}, where 
the black dotted curve is the inflationary trajectory 
on the ($\tau$, $\sigma$)-plane evaluated under 
the slow-roll approximation 
(\ref{eq:trajsigma}) and the red solid curve represents 
the same trajectory by solving the full equations of 
motion (\ref{eq:eqm}) numerically.


From the observational point of view, the above 
inflationary dynamics can be considered as a 
single-field inflation if the scalar 
density perturbation is successfully produced and 
in this case the inflation mechanism is essentially 
categorized into the so-called 
natural inflation~\cite{Freese:1990rb}. (With our 
parameter settings, the value of $\lambda$ defined 
in Eq.~(\ref{eq:deflambda}) is almost equal to $1$ due to 
$\sigma_{\rm inf} \sim \langle \sigma \rangle$ 
caused by the parameter choice~(\ref{parameter})). 
Therefore the effective potential for the 
canonically normalized field $\phi^1 \equiv k\tau$ 
is given by 
\begin{equation}
V_{\rm eff} = \Lambda^4 (1-\lambda \,(\hat{c}\,\phi^1)),
\label{eq:infpoeff}
\end{equation} 
where $\hat{c} \equiv c/k$ and 
$k\equiv \sqrt{2(K_{\rm eff})_{\hat{T}^1\bar{\hat{T}}^1}} =
\sqrt{\frac{(\langle T^2\rangle -b\,\sigma)^2 +2b^2\,\sigma^2}
{2\sigma^2 (\langle T^2\rangle -b\,\sigma)^2}}$
and the slow-roll parameters are explicitly 
shown in terms of $\phi^1$ as 
\begin{align}
\epsilon &= \frac{M_{\rm Pl}^2}{2}
\left(\frac{V^{'}}{V}\right)^2 
=\frac{(\hat{c}\,M_{\rm Pl})^2}{2} \lambda^2 
\frac{1-{\rm cos}^2 (\hat{c}\,\phi^1)}
{\left( 1-\lambda\,{\rm cos} (\hat{c}\,\phi^1)
\right)^2}, \nonumber\\
\eta &=M_{\rm Pl}^2
\frac{V^{''}}{V} 
=(\hat{c}\,M_{\rm Pl})^2 \lambda 
\frac{{\rm cos} (\hat{c}\,\phi^1)}
{1-\lambda\,{\rm cos} (\hat{c}\,\phi^1)}
\end{align}
those yield
\begin{align}
r&=16\,\epsilon, \nonumber\\
\xi^2 &=M_{\rm Pl}^4
\frac{V^{'} V^{'''}}{V^2} 
=-2(\hat{c}\,M_{\rm Pl})^2 \epsilon, 
\end{align}
where the prime denotes the derivative $d/d\phi^1$ 
with respect to the canonically normalized inflaton 
field $\phi^1$. 

The axion identified as the inflaton in the 
terminology of the natural inflation here 
corresponds to the zero mode of fifth component of 
the $U(1)_{I'=1}$ gauge field, $A_y^{I'=1}$, 
in our framework of 5D supergravity models, 
and here the axion decay constant $f_{\phi^1}$ is 
given by $\hat{c}^{-1}$. Although we need 
the large axion decay constant $f_{\phi^1} \geq M_{\rm Pl}$ 
in order to get the large tensor-to-scalar ratio 
$r\sim{\cal O}(0.1)$ in the natural inflation, 
this large axion decay constant is obtained from the 
small $U(1)_{I'=1}$ charge $c$ shown in Eq.~(\ref{parameter}) 
in our framework. 
In addition to the natural realization of the large axion decay 
constant, the $\eta$ problem peculiar to the general 
four-dimensional supergravity models is avoided here, 
because the K\"ahler potential does not 
include the axion field $\tau$ whose appearance is prohibited 
by the $U(1)_{I'=1}$ symmetry. 

In the same way as the case of small-field 
inflation, we denote the field values 
$(\sigma,\tau) =(\sigma_\ast,\tau_\ast)$ 
corresponding to the pivot scale, 
the number of e-foldings $N=N_\ast$ and 
the height of scalar potential $V=V_\ast 
\equiv V(\sigma_\ast,\tau_\ast)$ at the pivot scale 
as well as $V=V_{\rm end} \equiv V(\sigma_{\rm end}, \tau_{\rm end})$ 
at the end of inflation. In terms of them, 
the following e-foldings number 
$N_{\rm e} \equiv N_{\rm end}-N_\ast$ can be 
estimated as~\cite{Liddle:1993fq}
\begin{align}
N_{\rm e} \simeq 62 + 
\text{ln} \frac{V_{\ast}^{1/4}}{10^{16}\,{\rm GeV}} 
+\text{ln} \frac{V_\ast^{1/4}}{V_{\rm end}^{1/4}} 
-\frac{1}{3} \text{ln} \frac{V_{\rm end}^{1/4}}{\rho_R^{1/4}},
\label{eq:efold2real}
\end{align}
where we used $V_{\rm end}^{1/4} 
\simeq 1.9 \times 10^{16}\,{\rm GeV}$, 
$\rho_R^{1/4} =(\pi^2 g_\ast/30)T_R \simeq 
2.0 \times 10^{10}\,{\rm GeV}$. 
The effective degrees of freedom 
of the radiation $g_\ast=915/4$ 
at the reheating temperature $T_R$ 
can be fixed by assuming the MSSM with 
$T_R \simeq 6.8 \times 10^{9}\,{\rm GeV}$ 
whose numerical value will be determined 
later in Sec. \ref{subsec:rehl}. On the other hand, 
the same e-folding number $N_{\rm e}$ is also 
evaluated by  
\begin{align}
N_{\rm e} = -\int^{t_\ast}_{t_{\text{end}}} H(\tilde{t})\,d\tilde{t},
\label{eq:efold2ast}
\end{align}
therefore we find the numerical values $\sigma_{\ast}$, 
$ \tau_\ast$ and $V_{\ast}^{1/4}$ by equaling 
Eq. (\ref{eq:efold2real}) to Eq. (\ref{eq:efold2ast}),
\begin{equation}
\sigma_\ast \,\simeq\, 0.98,\qquad \tau_\ast \,\simeq\, 17.9, \qquad
V_\ast^{1/4} \,\simeq\, 1.9\times 10^{16}\,{\rm GeV}, 
\qquad N_{\ast}\,=\,16.6, 
\qquad N_{\rm e} \,=\, 64.6.
\end{equation}

\begin{figure}[h]
\centering \leavevmode
\includegraphics[width=0.5\linewidth]{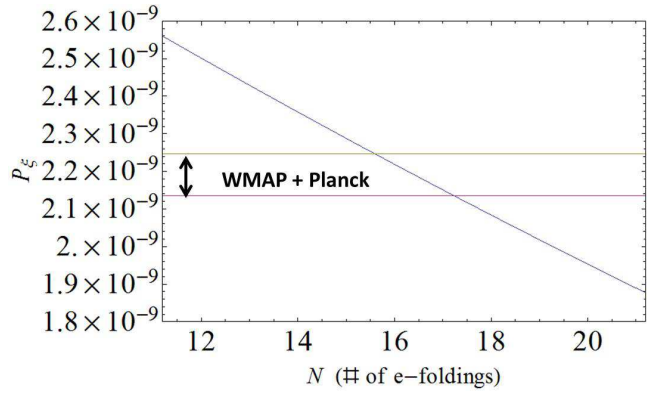}
\caption{The power spectrum of scalar curvature 
perturbation between 70 and 60 
e-foldings before the end of inflation.}
\label{fig:psl}
\end{figure}

\begin{figure}[h]
\centering \leavevmode
\includegraphics[width=0.5\linewidth]{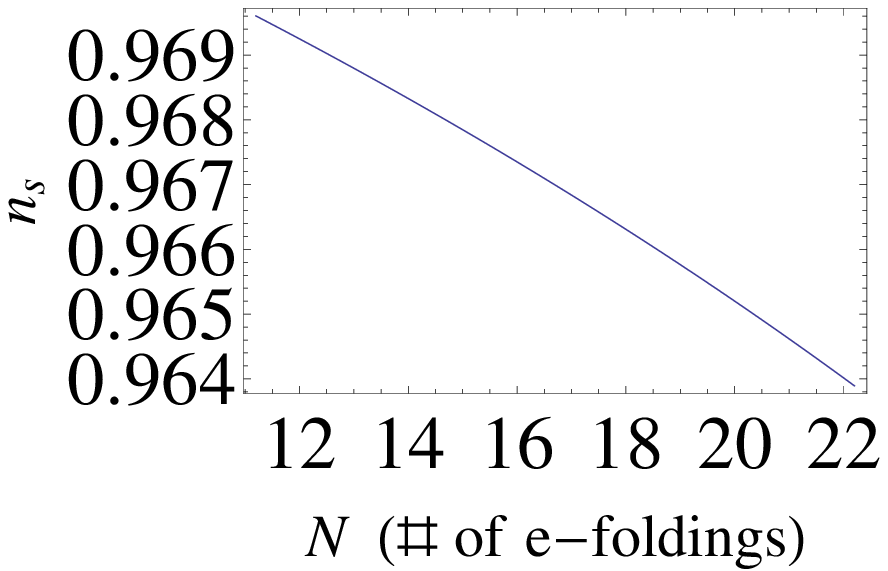}
\caption{The spectral index of scalar curvature 
perturbation between 70 and 60 
e-foldings before the end of inflation.}
\label{fig:sil}
\end{figure}

\begin{figure}
\centering \leavevmode
\includegraphics[width=0.5\linewidth]{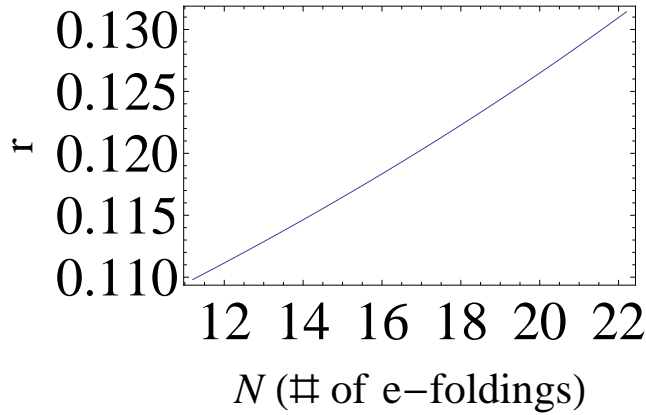}
\caption{The tensor-to-scalar ratio 
between 70 and 60 
e-foldings before the end of inflation.}
\label{fig:tts}
\end{figure}

Next, we check whether the power spectrum $P_\xi$ of scalar 
curvature perturbation, its spectral index $n_s$, the running 
of its spectral index $dn_s/d\,\ln k$ 
and the tensor-to-scalar ratio $r$, all at 
the pivot scale $k_0$, can be consistent with the 
recent observations or not. 
Especially, the BICEP2 collaboration~\cite{Ade:2014xna} has 
reported that a large value of 
the tensor-to-scalar ratio, 
\begin{equation}
r\,=\,0.16^{+0.06}_{-0.05},
\end{equation}
after considering the foreground dust. We extract 
the numerical values of these observables from our 
model as follows, 
\begin{equation}
 P_\xi \,=\,2.18 \times 10^{-9},\,\,\,
n_s \,=\, 0.967, \,\,\,
dn_s/d\ln k \,=\, -5.3\times 10^{-4}, \,\,\,
r\,=\,0.12,
\end{equation}
where the running of the spectral 
index is 
defined by ${\rm d}n_s/{\rm d}\,{\rm ln}\,k =-24\epsilon^2 
+16\epsilon\,\eta -2\xi^2$. 
These results of inflaton dynamics are summarized in 
Figs. \ref{fig:psl}, \ref{fig:sil} and \ref{fig:tts}. 
Note that our estimations are consistent with recent studies
~\cite{Freese:2014nla} reporting the consistency of the natural 
inflation with recent observations. 
These predictions are similar to those of the chaotic inflation, 
because the scalar potential (\ref{eq:infpoeff}) 
is similar to that of the chaotic inflation~\cite{Linde:1983gd} 
in the parameter region of the large axion decay constant 
$f_{\phi^1}=\hat{c}^{-1}\ge M_{\rm Pl}$. 
Note that this inflation mechanism is classified as the so-called 
large-field inflation whose change of the canonically normalized inflaton 
field $\phi^1$ is given by 
\begin{equation}
\Delta \phi^1 \,= \,\phi^1_\ast -\phi^1_{\rm end} 
\simeq 14.6\,M_{\rm Pl}.
\end{equation}
This kind of natural inflation scenario with the large 
axion decay constant is also discussed in Ref.~\cite{Kim:2004rp}, 
where the large axion decay constant is effectively generated 
from sub-Planckian decay constants.

In the following section \ref{subsec:modpro} 
and \ref{subsec:rehl}, we discuss the field oscillation 
during inflationary era and the moduli-induced 
gravitino problem via the inflaton decay and 
the reheating process.

\vspace{2cm}
\subsection{Moduli problem}
\label{subsec:modpro}
In this section, we consider the cosmological 
moduli problem~\cite{Lyth:1995ka} 
such as moduli-induced gravitino problem~\cite{Endo:2006zj} 
and the effect of field oscillation after the inflation. 
In our both inflation models, 
the moduli do not induce the SUSY breaking, 
therefore they do not decay into the gravitino 
which means that there is no moduli-induced 
gravitino problem. 
Even if there is a source of the SUSY breaking in 
the superpotential, the moduli will not get the F-term 
because they have large supersymmetric masses. 

In addition to the above issues, we have to check the field oscillation after 
inflationary era, because if the fields other than the inflaton 
oscillate after the inflation and dominate 
the universe, they affect the particle cosmology. 
Since both $\hat{T}^2$ and $H_2$ have a supersymmetric 
mass larger than the inflaton mass, we expect that these fields 
do not oscillate. 

By contrast, the stabilzer field $H_1$ and 
the inflaton get the same order of the supersymmetric 
mass as each other at the vacuum, and then $H_1$ could 
have been stabilized at a point different from its true 
minimum during the inflation 
and oscillated around the minimum after the inflation. 
In the inflationary era however, 
the mass of $H_1$ is given by the Hubble-induced mass 
proportional to $H_{\rm inf}$ shown in Eqs.~(\ref{eq:sfhinf}) 
and (\ref{eq:hubl}) in the small- and the large-field inflation 
scenarios proposed in the previous Sec.~\ref{sec:sinflation} 
and here in this Sec.~\ref{sec:linflation}, respectively. 

Therefore, in each scenario, $H_1$ is fixed strictly 
at the origin during and after inflation and does not oscillate 
and dominate the universe. 
We further remark that, in the case of small-field inflation 
discussed in Sec.~\ref{sec:sinflation}, 
the imaginary part of modulus Im\,$T^1$ does not oscillate 
as well if the initial position of Im\,$T^1$ is located at the 
origin. This is because 
the K\"ahler potential has a shift symmetry for the 
imaginary part and the inflationary dynamics does not 
involve the imaginary direction. 

In order to estimate the effects of supersymmetry 
breaking on the inflation dynamics, 
we consider the following superpotential,
\begin{align}
W &=W_{\rm eff} +\Delta W (\hat{T}^1), 
\label{eq:supbrea}
\end{align}
where $W_{\rm eff}$ represents the superpotential 
terms responsible for the inflation given in 
Eq.~(\ref{eq:weff3}), and 
$\Delta W(\hat{T}^1)$ describes the supersymmetry 
breaking sector, which may involve the inflaton 
multiplet $\hat{T}^1$ in general. 
We assume that the other 
fields such as $\hat{T}^2$ and $H_2$ 
are stabilized at their supersymmetric minimum. 
The following analysis can be applied to 
both the inflation scenarios by replacing 
the original effective superpotential $W_{\rm eff}$ 
with the modified one (\ref{eq:supbrea}) in each 
scenario.
The position of the supersymmetry breaking minimum 
will be determined by estimating the deviation from the 
supersymmetric Minkowski minimum (\ref{eq:movev}) 
by assuming 
$\langle \Delta W \rangle \sim 
\langle \partial_{\hat{T}^1}(\Delta W \rangle) \ll 1$ 
(in the unit $M_{\rm Pl}=1$) 
and employing the reference point method~\cite{Abe:2006xp}. 

As the reference point which should be 
selected as close to the true minimum as 
possible, 
we set it in such a way that the following conditions, 
\begin{align}
&D_{H_1} W|_{\rm ref} = W_{H_1} +(K_{\rm eff})_{H_1} W =0, \nonumber\\
& \hspace{2cm} \leftrightarrow c\,\hat{T}^1|_{\rm ref} 
=\text{ln} \frac{J_L^{(1)}}{J_0^{(1)}} \,\,{\rm and} 
\;\;H_1|_{\rm ref}=0, \nonumber\\
&D_{\hat{T}^1} W|_{\rm ref} = (K_{\rm eff})_{\hat{T}^1} \Delta W,
\end{align}
are satisfied at the point, where the effective K\"ahler 
potential $K_{\rm eff}$ is given in Eqs.~(\ref{eq:weff}) or 
(\ref{eq:weff3}) for each scenario. Then we expand the field 
$\varphi=\varphi |_{\rm ref} +\delta \varphi$ for 
$\varphi=\hat{T}^1, H_1$ 
and evaluate
the deviations $\delta \varphi$ from the reference point 
$\varphi |_{\rm ref}$. 
We find the following variations, 
\begin{align}
\delta \hat{T}^1 ={\cal O}\left( \frac{|\Delta W|^2}{W_{\hat{T}^1H_1}} \right),
\;\;\; \delta H_1= -\frac{(K_{\rm eff})_{\hat{T}^1} \Delta W}{W_{\hat{T}^1H_1}} +{\cal O}(|\Delta W|^2),
\end{align}
minimize the scalar potential at their first order, 
which implies our reference point method is 
valid if the supersymmetry breaking scale is smaller than 
the supersymmetric masses of the moduli and stabilizers, that is, 
$\langle \Delta W \rangle \ll \langle W_{\hat{T}^1H_1}\rangle$ 
in the unit $M_{\rm Pl}=1$. In the same way, 
the F-terms of $\hat{T}^1$ and $H_1$ at the supersymmetry 
breaking minimum are estimated as 
\begin{align}
\sqrt{(K_{\rm eff})_{\hat{T}^1\bar{\hat{T}}^1}} F^{\hat{T}^1} &= 
-e^{K_{\rm eff}/2} \sqrt{(K_{\rm eff})_{\hat{T}^1\bar{\hat{T}}^1}} 
(K_{\rm eff})^{\hat{T}^1\bar{J}}\,\overline{D_JW} 
\simeq {\cal O}\left(\frac{(m_{3/2})^3}{(m_{t^1})^2}\right),\nonumber\\
\sqrt{(K_{\rm eff})_{H_1\bar{H}_1}} F^{H_1} &= 
-e^{K_{\rm eff}/2} \sqrt{(K_{\rm eff})_{H_1\bar{H}_1}} 
(K_{\rm eff})^{H_1\bar{J}}\,\overline{D_JW} 
\simeq {\cal O}\left(\frac{(m_{3/2})^3}{(m_{h_1})^2}\right),
\end{align}
where $(m_{t^1})^2$ and $(m_{h_1})^2$ 
are given in Eq.~(\ref{eq:massd}). 

We conclude that 
if the size of supersymmetry breaking is much smaller 
than the inflation scale which we assume in this paper, 
they do not affect the 
inflation mechanism and the related cosmology after 
the inflation. In fact, the field $\hat{T}^1$ and $H_1$ 
have almost vanishing F-terms which means that 
the decay channels from $\hat{T}^1$ and $H_1$ 
into the gravitino are suppressed and they do not 
induce the moduli-induced gravitino problem. 
The coherent oscillation of $H_1$ after the inflation 
is also suppressed, because the amplitude of the 
oscillation of $H_1$, 
\begin{align}
&\Delta H_1\,\simeq \, \delta H_1|_{\rm inf} -
\delta H_1|_{\rm vac}\,\simeq\, 
{\cal O}\left( \frac{\Delta W}{H_{\rm inf}} \right) 
-{\cal O}\left( \frac{\Delta W}{m_{h_1}} \right),
\end{align}
is small enough, where $\delta H_1|_{\rm inf}$ and 
$\delta H_1|_{\rm vac}$ are the deviations of $H_1$ 
from the supersymmetric Minkowski minimum~(\ref{eq:movev}) 
during the inflation and at the true minimum where the 
supersymmetry is broken, respectively.

\subsection{Reheating temperature}
\label{subsec:rehl}
Finally, we show the decay channel and the reheating 
process after the end of inflation. 
As shown in Fig. \ref{fig:linftra}, after the inflation, 
both $\sigma$ and $\tau$ oscillate around the 
minimum and they decay into particles in the MSSM 
at the time 
$t_{\rm dec}^{t^1}$ and $t_{\rm dec}^{\phi^1}$ 
respectively where $t^1$ and $\phi^1$ are the 
canonically normalized field, 
$t^1\,=\,\sqrt{2(K_{\hat{T}})_1}U_{1,1}\sigma +
\sqrt{2(K_{\hat{T}})_1}U_{1,2}({\rm Re}\,\hat{T}^2)$, 
and $\phi^1 =\sqrt{2(K_{\rm eff})_{\hat{T}^1\bar{\hat{T}}^1}}\tau$ given in  Eqs.~(\ref{eq:canbas}) 
and (\ref{eq:canbasphi}), respectively. 
Note that the eigenvalue $(K_{\hat{T}})_{I'}$ 
and diagonalizing matrix $U_{I',J'}$ ($I'=1,2$) 
of the K\"ahler metric 
are explicitly shown in Appendix 
\ref{app:can}. In the following analysis, 
we neglect the oscillation of Re\,$\hat{T}^2$ 
and use the sudden-decay approximation.

The decay time $t_{\rm dec}^{t^1}\,=\,1/\Gamma^{t^1}$ 
is the inverse of the decay width of $t^1$ which depends 
on the concrete model of the particle physics. 
We assume that the modulus $t^1$ mainly 
decay into the gauge boson pairs, $t^1 \to g^{(a)}+g^{(a)}$, for simplicity. 
If some matter chiral multiplets $Q$ 
originating in the hypermultiplet ${\bm H}_\alpha$ 
have $U(1)_{I'=1,2}$ charges under the $Z_2$-odd 
vector multiplets $V^{I'=1,2}$ carrying the inflaton, 
they induce couplings like $Z(t^1)|Q|^2$ in 
the K\"ahler potential where 
$Z(t^1)$ is the K\"ahler metric of $Q$ given by 
Eq. (\ref{eq:km}) where the $U(1)_{I'=1,2}$ charges are 
replaced by those of $Q$. 
These couplings will enhance the inflaton decay width into $Q$ 
depending on the $U(1)_{I'=1,2}$ charge of $Q$. 
\footnote{If the $U(1)_{I'=1,2}$ charge of $Q$ is of 
${\cal O}(1)$, 
the decay width into $Q$ is of almost the same order as that 
of $\Gamma^{t^1} (t^1 \rightarrow g^{(a)}+g^{(a)})$.} 
In our model, the decay channel via the 
F-term of $t^1$ is kinematically forbidden, 
because its vacuum expectation value is negligibly small 
as mentioned previously. 
After all, $t^1$ mostly decays into the gauge boson pairs 
with the following decay width, 
\begin{align}
\sum_{a=1}^3
\Gamma^{t^1} (t^1 \rightarrow g^{(a)}+g^{(a)})&\simeq 
\sum_{a=1}^3\cfrac{N_G^a}{64\pi } 
\left\langle\cfrac{\xi_a^1}{{\rm Re}\,f_a}\right\rangle^2 
\left\langle
\frac{U_{2,2}}{\sqrt{2(K_{\hat{T}})_1}(U_{1,1}U_{2,2}-
U_{1,2}U_{2,1} )} \right\rangle^2 
\cfrac{m_{t^1}^3}{M_{\rm Pl}^2}
\nonumber\\
&\simeq 0.86\,\,{\rm GeV},
\label{eq:t1dec}
\end{align}
where $N_G^a$ is the number of the gauge bosons 
for the gauge group $G^a$ with $a=1,2,3$ representing 
the three gauge groups in the MSSM, $a=\,U(1)_Y$, $SU(2)_L$, 
$SU(3)_c$, respectively. We are adopting the numerical values 
of input parameters, those yield 
$\langle \sqrt{2(K_{\hat{T}})_1}\rangle \simeq 0.86$, 
$\langle U_{1,1}\rangle \simeq -39.68$, 
$\langle U_{2,1}\rangle \simeq 0.025$, 
$\langle U_{1,2}\rangle \,=\,
\langle U_{2,2}\rangle \,=\,1$ and 
$m_{t^1} \simeq 1.96\times 10^{13}\,{\rm GeV}$. 
Especially we set $\xi_a^1=3.58$ and $\xi_a^{I' \ne 1}=0$ 
to realize the correct gauge coupling 
$\langle f_a\rangle \,=\,1/(g_a)^2 \simeq 3.73$ 
at the grand unification scale 
($\simeq\,2.0\times10^{16} [\text{GeV}]$). 
Because 
$\Gamma^{t^1} (t^1 \rightarrow g^{(a)}+g^{(a)})$ 
with $a=1,2,3$ are assumed to be dominant, the total decay width is 
given by 
\begin{align}
\Gamma^{t^1} \simeq 
\sum_{a=1}^3
\Gamma^{t^1} (t^1 \rightarrow g^{(a)}+g^{(a)}).
\end{align}

On the other hand, the decay time $t_{\rm dec}^{\phi^1} 
\,=\,1/\Gamma^{\phi^1}$ 
is estimated from the following terms in the Lagrangian,
\begin{align}
{\cal L}&\supset -\cfrac{1}{8} 
{\rm Im}\,f_a\epsilon^{\mu\nu\rho\sigma} F^a_{\mu\nu}F^a_{\rho\sigma} 
\nonumber\\
&=-\cfrac{1}{8} 
\langle {\rm Im}\,f_a\rangle\epsilon^{\mu\nu\rho\sigma} 
F^a_{\mu\nu}F^a_{\rho\sigma} 
-\cfrac{1}{8}\left\langle\cfrac{\partial\,{\rm Im}\,f_a}{\partial \phi^1}\right\rangle 
\delta \phi^1 \,\epsilon^{\mu\nu\rho\sigma} F^a_{\mu\nu}F^a_{\rho\sigma}.
\end{align}
Then the total decay width of the field $\phi$ is computed as follows,
\begin{align}
\Gamma^{\phi^1}\simeq
\sum_{a=1}^3
\Gamma^{\phi^1} (\phi^1 \rightarrow g^{(a)}+g^{(a)})&\simeq 
\sum_{a=1}^3\cfrac{N_G^a}{128\pi } 
\left\langle\cfrac{\xi_a^1}{\sqrt{(K_{\rm eff})_{\hat{T}^1\bar{\hat{T}}^1}}{\rm Re}\,f_a}\right\rangle^2 
\cfrac{m_{\phi^1}^3}{M_{\rm Pl}^2}
\nonumber\\
&\simeq 1359\,\,{\rm GeV},
\label{eq:phidec}
\end{align}
where the given input parameters lead to 
$\sqrt{(K_{\rm eff})_{\hat{T}^1\bar{\hat{T}}^1}} \simeq 0.61$ and 
$m_{\phi^1} \simeq 1.96\times 10^{13}\,{\rm GeV}$. 
Since both the fields $t^1$ and $\phi^1$ have the almost degenerate 
supersymmetric masses (\ref{eq:massvac}), 
the differences 
between $\Gamma^{t^1}$ shown in Eq.~(\ref{eq:t1dec}) 
and $\Gamma^{\phi^1}$ in Eq.~(\ref{eq:phidec}) come from the 
K\"ahler metric when the fields $\sigma$ and $\tau$ are 
diagonalized. 

From the expressions (\ref{eq:t1dec}) and (\ref{eq:phidec}), 
we find the decay time $t^{\phi^1}$ is much smaller than $t^{t^1}$, 
i.e., $t^{\phi^1} \ll t^{t^1}$. It indicates that the inflaton $\phi^1$ 
decay into the radiation faster than the decay of 
the real part of the modulus $t^1$ into the radiation. 
Then the reheating temperature is estimated by 
equaling the expansion rate of the universe and the total decay width, 
\begin{align}
\Gamma^{\phi^1}&\simeq H (T_R), \nonumber\\
\Leftrightarrow T_{R} &= 
\left( \cfrac{\pi^2 g_\ast}{90}\right)^{-1/4} 
\sqrt{\Gamma^{\phi^1} M_{\rm Pl}} \simeq  6.8\times 10^9\,{\rm GeV},
\end{align}
where $g_\ast =915/4$ is the effective degrees of 
freedom of the radiation at the reheating in the MSSM. 

Since $t^1$ behaves as the non-relativistic 
particle after the inflation, its energy density decreases as $a^{-3}$ 
compared to that of the radiation $a^{-4}$, where $a$ is the scale factor. 
Thus whether there is a second  reheating or not after $t^1$ 
decays depends on the following condition. If the following 
condition is satisfied, $t^1$ dominates the universe and 
it induces the second reheating,
\begin{equation}
1 \leq \frac{\rho_{t^1}}{\rho_R}\biggl|_{T=T_{t^1}} \simeq 
\frac{\rho_{t^1}}{\rho_R}\biggl|_{T=T_R} 
\left( \frac{T_{R}}{T_{t^1}} \right),
\label{eq:dom}
\end{equation}
where $\rho_{t^1}$ and $\rho_R$ are the energy densities 
of $t^1$ and the radiation, respectively, and 
$T_{t^1}$ is the decay temperature of $t^1$ given by 
\begin{align} 
T_{t^1} &= 
\left( \cfrac{\pi^2 g_\ast}{90}\right)^{-1/4} 
\sqrt{\Gamma^{t^1} M_{\rm Pl}} \simeq  1.7\times 10^8\,
{\rm GeV}.
\end{align}
After the inflation, the field $t^1$ and the inflaton $\phi^1$ 
oscillate at the same time and the difference between 
them is only the size of the decay 
width. Thus we expect that the amplitude of 
$\Delta t^1$ is small enough at the time $t^{\phi^1}$ 
and the energy density $\rho_{t^1}\simeq 
 m_{t^1}^2(\Delta t^1)^2$ is neglected compared 
to that of the radiation 
$\rho_R^{1/4}\,=\,2\times 10^{10}\,{\rm GeV}$. 
It follows that the above condition~(\ref{eq:dom}) 
is not satisfied, and then the second reheating does 
not occur. 

Finally we comment on the one-loop corrections to the 
moduli K\"ahler potential given by Eq. (\ref{eq:oneloop}). 
Although the loop correction to the effective K\"ahler 
potential depends on the moduli Re\,$\hat{T}^{I'}$ with $I'=1,2$ 
via the Norm function ${\cal N}$ shown in Eq~(\ref{eq:norm}), 
the Re\,$\hat{T}^1$-dependence of the potential is similar to that 
of the tree-level K\"ahler potential. Since the scalar potential 
also diverges in the limit $b\,{\rm Re}\,\hat{T}^1 
\rightarrow {\rm Re}\,\hat{T}^2$ due to the behavior of the 
following factor in this limit, 
\begin{equation}
e^{K_{\rm eff}} \simeq \frac{e^{1/(32\pi^2 {\cal N})}}{{\cal N}} 
\rightarrow \infty,
\end{equation}
the modulus ${\rm Re}\,\hat{T}^1$ is not destabilized during 
and after the inflation. 
Such a behavior implies that the field Re\,$\hat{T}^1$ 
remains stabilized during the inflation, which is 
considered as the single-field inflation 
with the imaginary part of modulus identified as the inflaton. 

\section{Conclusion}
\label{sec:conclusion}
In this paper, we proposed the effective mechanism 
to realize successful inflation to explain the  
cosmological observations based on the 5D 
supergravity models on $S^1/Z_2$. 
In our framework, we can realize both the small- 
and the large-field inflation scenarios, where 
the role of inflaton is played by a linear combination 
of the moduli appearing after 
compactifying the fifth direction. 
These two inflation scenarios would be compatible 
with numerous particle physics models constructed in 
5D. 

In the case of the small-field inflation, 
the real part of the light modulus, Re\,$T^1$, 
is considered as the inflaton and the inflaton potential is induced 
by the superpotential of the stabilizer field, $H_1$, 
which has a localized wavefunction in the fifth 
dimension. 
This small-field inflaton potential is consistent with 
WMAP and Planck data~\cite{Ade:2013zuv}, 
although they cannot explain the large tensor-to-scalar 
ratio reported by BICEP2~\cite{Ade:2014xna}. 
We have also studied the particle cosmology in this 
inflationary scenario and find that there is no moduli 
and gravitino over production. 

We further presented a different setup within the 
same 5D supergravity framework, realizing the 
large-field inflation which produces the sizable 
tensor-to-scalar ratio consistent with the results 
from BICEP2. 
In this scenario, the two light pairs of moduli and 
stabilizer fields ($\hat{T}^{I'},H_i$) with $I',i=1,2$ are introduced 
and the inflaton is identified 
as the imaginary part of the lightest modulus Im\,$\hat{T}^1$. 
The moduli potential is induced by the superpotential of the 
stabilizer fields as in the small-field scenario. 
The inflaton potential is similar to the one of 
natural inflation~\cite{Freese:1990rb}, 
but in our framework, the axion decay constant 
is given by the $U(1)$ charges originated 
from the $Z_2$-odd vector multiplets carrying the inflaton field. 
Both the inflation scenarios proposed in this paper are 
free from the $\eta$ problem which is peculiar to the 
inflationary dynamics in the four-dimensional 
${\cal N}=1$ supergravity models. 

In the large-field scenario, when the imaginary part of the modulus 
rolls down in its potential, the real part of the modulus 
will be destabilized because there is a runaway 
direction in its potential in general.
However, in our model, the real part of the modulus, 
Re\,$\hat{T}^1$, can be stabilized during the inflation, because of the 
potential barrier is produced by the real 
part of the heavier modulus, Re\,$\hat{T}^2$, in the K\"ahler potential. 
After the inflation, 
both the imaginary and the real part of the 
modulus, Im$\,\hat{T}^1$ and Re$\,\hat{T}^1$ 
oscillate but only the inflaton Im\,$\hat{T}^1$ reheats the universe. 
The reheating temperature is estimated 
from the decay width of the inflaton 
into the gauge boson pairs. 
The stabilizer fields are also fixed at the origin 
during (-and after-) the inflation by the 
Hubble-induced and their own supersymmetric 
masses. Therefore there is no cosmological moduli 
problem also in the large-field scenario. 

Both the proposed inflation scenarios are insensitive 
to the supersymmetry breaking required by the particle 
phenomenology, if the breaking scale is lower than the 
inflation scale. 
This is because the large supersymmetric 
masses are provided from the superpotential 
of charged stabilizer fields, those are controlled 
by the $U(1)$ charges under the 5D vector 
multiplets carrying the moduli. 
The branching ratio of the moduli decaying 
into gravitino is suppressed due to such the 
supersymmetric masses of moduli. 
The field in the supersymmetry breaking sector 
may oscillate after the inflation if the size 
of supersymmetry breaking is smaller than 
the inflation scale. 
The further model building of the particle 
cosmology including the concrete matter 
sectors remains as a future work. 

The moduli potential as well as their kinetic terms 
are strictly constrained by the symmetries in 
higher-dimensional spacetime, although the moduli 
behave as Lorentz scalars in the four-dimensional spacetime 
with the extra-dimensions compactified. 
For the inflationary dynamics proposed in this paper, 
the $U(1)_{I'}$ symmetries played essential roles, 
those generate the localized wavefunctions of charged 
stabilizer zero-modes and then yield the suitable moduli 
potential. It would be possible that the 5D supergravity 
studied in this paper is derived as the 5D effective theory 
of supergravities in more-than-five dimensional spacetime, 
superstrings in ten-dimensions and the M-theory 
in eleven-dimensions~\cite{Lukas:1998yy}. 
In such cases, the coefficients $C_{I,J,K}$ in the norm function 
will be related to the geometric structure of the internal space 
(e.g., the intersection numbers of Calabi-Yau manifold) and the 
above $U(1)_{I'}$ symmetries might originate from certain local 
symmetries with the gauge fields in the higher-dimensional spacetime. 
Our 5D models not only work well observationally, but also would 
be theoretically instructive and extensible from the above points of view.

\subsection*{Acknowledgement}
The authors would like to thank T.~Higaki, Y.~Sakamura 
and Y.~Yamada for useful discussions and comments. 
The work of H.~A. was supported in part by the Grant-in-Aid for 
Scientific Research No.~25800158 from the Ministry of Education, 
Culture, Sports, Science and Technology (MEXT) in Japan. 
H.~O. was supported in part by a Grant-in-Aid for JSPS Fellows 
No. 26-7296 and a Grant for 
Excellent Graduate Schools from the MEXT in Japan.

\appendix
\section{The canonical normalization in the large-field model}
\label{app:can}
As we have seen in Sec. \ref{sec:linflation}, the 
moduli stabilization mechanism involves sizable moduli mixings 
in the K\"ahler metric and we have to canonically normalize 
the moduli to estimate their masses. 
In this appendix, we show the eigenvalue $(K_{\hat{T}})_{I'}$ 
and the diagonalizing matrix $U_{I'\bar{J}'}$ of the 
K\"ahler metric $K_{I'\bar{J}'}$.

The K\"ahler metric in the effective K\"ahler potential 
(\ref{eq:weff2}) derived from the norm function (\ref{eq:norm}) with $a=1$ is given by 
\begin{align}
(K_{\rm eff})_{I',\bar{J}'}\,=\,
\begin{pmatrix}
\left(\frac{1}{2\sigma^1} \right)^2 
+\frac{1}{2}\left(\frac{b}{\sigma^2-b\,\sigma^1}\right)^2 
& -\frac{b}{2\left(\sigma^2-b\,\sigma^1 
\right)^2}\\
-\frac{b}{2\left(\sigma^2-b\,\sigma^1 
\right)^2}
& \frac{1}{2\left(\sigma^2-b\,\sigma^1 
\right)^2}.
\label{app:kahmet}
\end{pmatrix}
,
\end{align}
where we define $\sigma^{I'}\equiv{\rm Re}\,\hat{T}^{I'}$ 
with $I'=1,2$ in this appendix. 
Then the eigenvalues of K\"ahler metric 
(\ref{app:kahmet}) are estimated as 
\begin{align}
(K_{\hat{T}})_1\,&=\,
\frac{(2+3b^2)(\sigma^1)^2 -2b\,\sigma^1\sigma^2 
+(\sigma^2)^2}{8(\sigma^1)^2 (\sigma^2-b\,\sigma^1)^2} 
+\frac{\sqrt{g(\sigma^1,\sigma^2)}}
{8(\sigma^1 )^2 
(\sigma^2-b\,\sigma^1 )^2},
\nonumber\\
(K_{\hat{T}})_2\,&=\,
\frac{(2+3b^2)(\sigma^1)^2 -2b\,\sigma^1\sigma^2 
+(\sigma^2)^2}{8(\sigma^1)^2 (\sigma^2-b\,\sigma^1)^2} 
-\frac{\sqrt{g(\sigma^1,\sigma^2)}}
{8(\sigma^1 )^2 
(\sigma^2-b\,\sigma^1 )^2},
\nonumber\\
g(\sigma^1,\sigma^2) &\equiv 
(4+4b^2+9b^4)(\sigma^1)^4 
+4b(2-3b^2)(\sigma^1)^3\sigma^2 
\nonumber\\
&\hspace{16pt}+2(5b^2-2)(\sigma^1)^2 (\sigma^2)^2 
-4b\,\sigma^1\,(\sigma^2)^3 +(\sigma^2 )^4.
\end{align}
The diagonalizing matrix of the K\"ahler metric 
(\ref{app:kahmet}) is given by
\begin{align}
U\,&=\,
\begin{pmatrix}
U_{1,1} & 1 \\
U_{2,1}& 1
\end{pmatrix}
,
\nonumber\\
U_{1,1}&=
\frac{(2-3b^2)(\sigma^1)^2 +2b\,\sigma^1\sigma^2 
-(\sigma^2)^2}{4b\,(\sigma^1)^2} 
-\frac{\sqrt{g(\sigma^1,\sigma^2)}}
{4b\,(\sigma^1 )^2},
\nonumber\\
U_{2,1}&=
\frac{(2-3b^2)(\sigma^1)^2 +2b\,\sigma^1\sigma^2 
-(\sigma^2)^2}{4b\,(\sigma^1)^2} 
+\frac{\sqrt{g(\sigma^1,\sigma^2)}}
{4b\,(\sigma^1 )^2}.
\end{align}


\end{document}